\documentclass[a4paper,fleqn,usenatbib,useAMS]{mnras}

\usepackage{graphicx}	
\usepackage{amsmath}	
\usepackage{amssymb}	
\usepackage{multicol}        
\usepackage{bm}		
\usepackage{pdflscape}	

\usepackage[T1]{fontenc}
\usepackage{ae,aecompl}

\usepackage{txfonts}

\usepackage{epsfig}
\usepackage{breqn}
\usepackage{footmisc}

\title[Age--dating LRGs observed with SALT]{Age--dating Luminous Red Galaxies observed with the Southern African Large Telescope{\thanks{based on observations made with the Southern African Large Telescope (SALT).}}}

\author[Ratsimbazafy et al.]{
A. L. Ratsimbazafy$^{1,2}$\thanks{Contact e-mail:\href{mailto:Ando.Ratsimbazafy@nwu.ac.za}{Ando.Ratsimbazafy@nwu.ac.za}}, 
S. I. Loubser$^{1}$,
S. M. Crawford$^{3}$, 
C. M. Cress$^{4,2}$,
B. A. Bassett$^{5,3,6}$, 
\newauthor
R. C. Nichol$^{7}$,
and P. V\"ais\"anen$^{3,8}$\\
$^{1}$Centre for Space Research, North-West University, Potchefstroom 2520, South Africa\\
$^{2}$Physics Department, University of the Western Cape, Private Bag X17 Cape Town 7535, South Africa\\
$^{3}$South African Astronomical Observatory, PO Box 9 Observatory 7935, Cape Town, South Africa\\
$^{4}$Centre of High Performance Computing, CSIR, 15 Lower hope St., Rosebank, Cape Town 7700, South Africa\\
$^{5}$African Institute for Mathematical Sciences, 6-8 Melrose	 Road, Muizenberg, Cape Town 7945, South Africa\\
$^{6}$Department of Maths and Applied Maths, University of Cape Town, Rondebosch 7701, South Africa\\
$^{7}$Institute of Cosmology and Gravitation, Dennis Sciama Building, Burnaby road, Portsmouth, PO1 3FX, UK\\
$^{8}$Southern African Large Telescope, PO Box 9 Observatory 7935, Cape Town, South Africa }
\date{Accepted - . Received - ; in original form -}

\pubyear{2016}
\begin{document}
\label{firstpage}
\pagerange{\pageref{firstpage}--\pageref{lastpage}}

\maketitle

\begin{abstract}
We  measure a value for the cosmic expansion of $H(z) = 89 \pm 23$(stat) $\pm$ 44(syst) km s$^{-1}$ Mpc$^{-1}$ at a redshift of $z \simeq 0.47$ based on the differential age technique.  This technique, also known as cosmic chronometers, uses the age difference between two redshifts for a passively evolving population of galaxies to calculate the expansion rate of the Universe.  Our measurement is based on the analysis of high quality spectra of luminous red galaxies obtained with the Southern African Large Telescope in two narrow redshift ranges of $z \simeq 0.40$ and $0.55$ as part of an initial pilot study. Ages were estimated by fitting single stellar population models to the observed spectra.  This measurement presents one of the best estimates of $H(z)$ via this method at $z\sim0.5$ to date. 

\end{abstract}




\section{Introduction} \label{section:intro}

In recent times, numerous observations and methods have been used and proposed for constraining cosmological parameters. These include observations of the cosmic microwave background, from the {\it Wilkinson Microwave Anisotropy Probe\/} team \citep{Spergel03, Spergel07, Komatsu09, Komatsu11, Hinshaw13} and  {\it Planck\/} team \citep{Planck Collaboration 2014,Planck Collaboration 2015}, baryonic acoustic oscillations peaks \citep{Eisenstein05,Percival10, Anderson12}, Type Ia supernovae \citep[SNe;][]{Riess98, Perlmutter99}, gravitational lensing \citep{Refregier03,Cao12} and the abundance of clusters of galaxies \citep{Haiman01}. Many of these methods use quantities that require the Hubble parameter to be integrated along the line of sight (e.g. the luminosity distance in SNe observations) and thus probe average expansion over a long period. A complementary technique known as  \textit{cosmic chronometers} (CC), originally proposed by \citet{Jimenez02}, uses passively evolving early--type galaxies to measure the Hubble parameter at a specific redshift, rather than a quantity that gives an integrated measurement of $H(z)$ out to redshift $z$. This technique is independent of the cosmological model and allows a determination of the expansion rate at a given redshift without relying on the nature of the metric between the chronometers and observers, and therefore can provide tighter constraints on cosmological parameters of the model.

The expansion rate $H(z)$ at redshifts $z >$ 0 can be obtained by
\begin{equation}
H(z) = - \dfrac{1}{(1+z)} \dfrac{dz}{dt}.
\label{hzeqn}
\end{equation}
In the CC method, $dz/dt$ is approximated by determining the time interval $\varDelta t$ corresponding to a given $\varDelta z$, where $\varDelta z$ is centred at redshift $z$. If one assumes that most stars in luminous red galaxies (LRGs) formed near the beginning of the Universe at a similar time (as supported by observations cited in \citet{Jimenez02,Simon05,Stern10a,Carson10,Moresco12,Moresco15,Liu12,Zhang14}), then measuring the age difference between ensembles of LRGs at two different redshifts provides the differential quantity $\varDelta z/\varDelta t$ required to estimate $H(z)$.

The biggest advantage of this method is the differential approach where the relative ages $\varDelta t$ are used instead of the absolute ages. The relatives ages diminish the systematic effects underlying the absolute age estimates. Furthermore, the use of the small redshift slices $\varDelta z$, which corresponds to a short time evolution of these massive and passive galaxies, controls the effect of some observation biases such as the progenitor bias. More discussions about avoiding this effect can be found in \citet{Moresco16}. 

The CC method has been used to measure the Hubble parameter up to redshift $z\sim 2$ and these results have been used to constrain the nature of dark energy and to recover the local Hubble constant \citep{Ferreras01,Jimenez02,Ferreras03,Jimenez03,Capozziello04,Simon05,Verkhodanov05,Dantas07,Samushia10,Stern10a,Moresco11,Moresco12,Liu12,Zhang14,Moresco15,Moresco16}
. In each measurement, the authors assumed that LRGs are massive, passively-evolving elliptical galaxies that are homogeneous populations forming their stars at high redshift, and they fit single--burst equivalent ages to the galaxies. These galaxies represent the ideal population to trace the differential age evolution of the Universe. In many of the papers \citep[e.g.][]{Simon05,Carson10,Crawford10a,Stern10a}, the authors have attempted to improve this method by pointing out the need for better fitting of stellar population models, better selection of targets, larger samples, and better quality data to precisely determine $H(z)$.

The age--dating of galaxies is an important topic in the study of the galaxy evolution. During the last few decades, it has been widely exploited by the creation of synthetic stellar population tools. There are number of stellar population synthesis codes available 
\citep[e.g.][]{Buzzoni89,Worthey94,FiocRocca97,Leitherer99,BC03,LeBorgne04,Jimenez04,Gonzalez05,Maraston05,Tantalo05,Coelho07,Schiavon07,Conroy09,Elbridge09,Kotulla09,Molla09,Vazdekis10,Maraston11,Vazdekis12}  that can be used to generate synthetic spectra to determine different parameters such as age, metallicity, chemical abundance and star formation history of a galaxy. Typically, two approaches are used for fitting models to data: (i) full--spectrum fitting \citep[e.g.][]{Heavens00,CidFernandes05,Ocvirk06a,Ocvirk06b,Chilingarian07,Tojeiro07,Koleva09} and (ii) techniques that fit features in the spectra such as Lick--index fitting \citep[e.g.][]{Worthey97,BC03,Thomas03,Lee05,Annibali07,Schiavon07,Lee09,Vazdekis10,Thomas11}. We explored one of these techniques which is the full--spectrum fitting, as further outlined in Section \ref{section:analysis}. 

\citet{Crawford10a} found that a measurement of $H(z)$ within a 3\% accuracy would be viable from a large redshift programme targeting LRGs, and explored optimal observational set--ups for the experiment using the Southern African Large Telescope (SALT). The potential to obtain wide optical wavelength coverage and high--resolution spectra from a 10--m--class telescope like SALT provides an excellent opportunity to explore the potential of the CC method.
In this paper, we describe the results of several observations that use the Robert Stobie Spectrograph (RSS) on SALT to obtain 16 spectra of massive LRGs at $z \simeq 0.40$ and $0.55$. We age--date the LRGs and hence determine $H(z \simeq 0.47$). 

The paper is organized as follows: We first describe the LRG sample selection from the 2dF--SDSS LRGs catalogue (Section \ref{section:selectionandreduction}). We then describe the data reduction and some details about the method adopted to estimate the ages of these galaxies (Sections \ref{section:selectionandreduction} and \ref{section:analysis}). In Section \ref{section:hz}, we discuss the derived ages and estimate $H(z=0.47)$. 
\begin{table*}
 \centering
 \begin{minipage}{140mm}
  \caption{Characteristics of the galaxies observed with the SALT telescope. All galaxies were selected from the 2dF--SDSS LRG catalogue \citep{Cannon06}, except for SDSS J013403.82$+$004358.8 that was extracted from the photo--$z$ catalogue known as MegaZ--LRG \citep{Collister07}. Its spectroscopic redshift was taken by matching it with the cluster catalogue of \citet{Wen12}, and its magnitude in $g$--band was from NED data base$^\star$; however, all remaining magnitudes, i.e. in $V$--band were from SIMBAD data base. The galaxy redshifts were taken from the catalogue. The $E(B-V)$ values were obtained using the reddening maps of \citet{Schlegel98} and the extinction curve of the \citet{Fitzpatrick99} with $R=3.1$.} 
  \begin{tabular}{lccccccl}
  \hline
   Name    &   RA         & Dec.    & $V$ & Redshift & log$M_{*}$& $E(B-V)\rm_{galactic}$ \\
               & (2000.0) & (2000.0)&(mag)&           & $(\rm M_{\sun})$& (mag)\\
 \hline
 2SLAQ J081258.12$-$000213.8&08 12 58.1&$-$00 02 14&19.65&0.4063&11.60$\pm$0.05&0.039\\
 2SLAQ J081332.20$-$004255.1&08 13 32.2&$-$00 42 55&19.62&0.3924&11.62$\pm$0.11&0.031\\
 2SLAQ J100315.23$-$001519.2&10 03 15.2&$-$00 15 19&19.62&0.3980&11.29$\pm$0.06&0.039\\
 2SLAQ J100825.72$-$002443.3&10 08 25.7&$-$00 24 43&19.44&0.3984&11.41$\pm$0.06&0.033\\
 2SLAQ J134023.93$-$003126.8&13 40 23.9&$-$00 31 27&18.89&0.3997&11.46$\pm$0.05&0.032\\
 2SLAQ J134058.83$-$003633.6&13 40 58.8&$-$00 36 34&19.29&0.4097&11.38$\pm$0.06&0.026\\
 2SLAQ J092612.79$+$000455.8&09 26 12.8&$+$00 04 56&20.52&0.5411&11.64$\pm$0.11&0.032\\
 2SLAQ J092740.75$+$003634.1&09 27 40.7&$+$00 36 34&20.25&0.5480&11.40$\pm$0.06&0.038\\
 2SLAQ J100121.88$+$002636.4&10 01 21.9&$+$00 26 36&19.48&0.5549&11.25$\pm$0.08&0.025\\
 2SLAQ J100131.77$-$000548.0&10 01 31.8&$-$00 05 48&20.03&0.5464&11.51$\pm$0.06&0.035\\
 2SLAQ J104118.06$+$001922.3&10 41 18.0&$+$00 19 22&20.02&0.5465&11.52$\pm$0.07&0.055\\
 2SLAQ J144110.62$-$002754.5&14 41 10.6&$-$00 27 54&20.55&0.4003&11.76$\pm$0.05&0.038\\
 2SLAQ J010427.15$+$001921.5&01 04 27.1&$+$00 19 21&19.71&0.4071&11.52$\pm$0.06&0.035\\
 2SLAQ J022112.71$+$001240.3&02 21 12.7&$+$00 12 40&20.80&0.3975&11.15$\pm$0.06&0.036\\
 2SLAQ J225540.39$-$001810.7&22 55 40.4&$-$00 18 11&20.10&0.5512&11.69$\pm$0.06&0.086\\
 SDSS J013403.82$+$004358.8&01 34 03.8&$+$00 43 59&20.3g&0.4092&11.96$\pm$0.13&0.024\\
\hline
\multicolumn{7}{l}{$^{\star}$\url{http://ned.ipac.caltech.edu/}}
\label{tab:tabchar}
\end{tabular}
\end{minipage}
\end{table*}
\begin{table*}
 \centering
 \begin{minipage}{180mm}
  \caption{The observing log. The first 10 galaxies were observed during the semester 2011-3 run, and the last 4 during the first semester of 2012-1 observations. }
  \begin{tabular}{lcccccccl}
  \hline
  Name &  Date   &  Exposure time& Slit & Grism & Resolution& PA & Comments\\
   
        & YYYY-MM-DD & (s)&(arcsec)&(grooves mm$^{-1}$ )&(\AA{} pix$^{-1}$) & ($^{\circ}$) & moon,seeing\\
  
 \hline
 2SLAQ J081258.12$-$000213.8&2012-01-25&2883&1&900&0.95&~~~0.0&dark grey, 1.5 arcsec\\
 2SLAQ J081332.20$-$004255.1&2011-12-20&2883&1&900&0.95&~~~0.0&dark grey, 1 arcsec\\
 2SLAQ J100315.23$-$001519.2&2012-01-29&2883&1&900&0.95&~~~0.0&dark grey, 1 arcsec\\
 2SLAQ J100825.72$-$002443.3&2012-01-22&2883&1&900&0.95&~~~0.0&dark grey, 1.4 arcsec\\
 2SLAQ J134023.93$-$003126.8&2012-02-21&2697&1&900&0.95&~~~0.0&dark, 1.95 arcsec\\
 2SLAQ J134058.83$-$003633.6&2012-02-20&2697&1&900&0.95&~~~0.0&dark, 1.3 arcsec\\
 2SLAQ J092612.79$+$000455.8&2012-01-26&2883&1&900&0.95&~~~0.0&dark grey, 1.4 arcsec\\
 2SLAQ J092740.75$+$003634.1&2012-01-26&2883&1&900&0.95&~~~0.0&dark grey, 1.4 arcsec\\
 2SLAQ J100121.88$+$002636.4&2012-02-28&3183&1&900&0.95&~~~0.0&grey, 1.6 arcsec, no flats\\
 2SLAQ J100131.77$-$000548.0&2012-02-20&3483&1&900&0.95&~~~0.0&grey, 1.3 arcsec\\
 2SLAQ J104118.06$+$001922.3&2012-02-23&3183&1&900&0.95&~~~0.0&grey, 1.7 arcsec\\
 2SLAQ J144110.62$-$002754.5&2012-07-19&3915 &1.5&900&0.95&~~~0.0&dark, 2.0 arcsec\\
 2SLAQ J010427.15$+$001921.5&2012-09-17&3714&1.5&900&0.95&~~69.5&dark, 1.5 arcsec\\
 2SLAQ J022112.71$+$001240.3&2012-09-19&3714&1.5&900&0.96&~115.0&dark, 1.46 arcsec\\
 2SLAQ J225540.39$-$001810.7&2012-09-20&3714&1.5&900&0.95&-161.0&dark, 1.5 arcsec\\
 SDSS J013403.82$+$004358.8&2012-09-20&3714&1.5&900&0.95&-132.0&dark, 1.9 arcsec\\
 
\hline
\label{tab:obslog}
\end{tabular}
\end{minipage}
\end{table*}

\section{Sample Selection and Data Reduction} \label{section:selectionandreduction}

\subsection{Sample Selection}

 Our targets are LRGs from 2dF--SDSS LRG and QSO (2SLAQ) catalogue\footnote{The 2SLAQ survey was a spectroscopy follow-up on the LRG targets based on SDSS DR4 photometric survey, focusing on targets beyond $z \geq 0.4$.} \citep{Cannon06} where their spectroscopic redshifts are already available, and from MegaZ--LRG photometric redshift catalogue\footnote{MegaZ--LRG catalogue was based on the SDSS DR5 photometric survey using the same selection criteria for the 2SLAQ LRG survey.} \citep{Collister07}. Those galaxies must be visible with SALT, massive, without emission lines, the brightest and the reddest in the catalogue. Detailed description on each criterion is described below. Many galaxies meet the selection criteria, but for a pilot survey, only few galaxies including those 16 galaxies listed in Table \ref{tab:tabchar} were aligned for the observations.

We estimated the stellar masses of LRGs using the method outlined below, requiring potential targets to have stellar masses larger than $10^{11}$ M$\rm_{\sun}$. We also required targets to have redshifts within 0.01 of $z = 0.40$ or $0.55$, and to have magnitude $V > 21$.

We derived the stellar masses of LRGs in 2SLAQ by first cross--matching the catalogue with the \textit{WISE}\footnote{\it{Wide--field Infrared Survey Explorer}} data and detected 13518 sources in at least two of the four mid--infrared bands centred at 3.4 (W1) and 4.6 $\mu$m (W2). The details on cross--matching the catalogue are provided in Appendix \ref{appendix:sedfitting}. The use of the \textit{WISE} data helped us to achieve our goal of selecting the most massive LRGs that have typically formed at high redshifts. Moreover, the combination of SDSS and \textit{WISE} photometry has been investigated to produce a new technique to target LRGs at high redshift ($z >$ 0.6). We then combined the optical and mid--infrared flux densities to build the spectral energy distribution (SED) of these detected sources. Next, we performed SED fitting with \textsc{cigale}\footnote{Code Investigating GAlaxy Evolution (\url{http://cigale.oamp.fr/})} \citep{Burgarella05}. Details of the mechanisms of \textsc{cigale} can be found in \citet{Noll09} and \citet{Giovannoli11}. We adopted basic input parameters as suggested by the code developer (Giovannoli, private communication) to extract the derived stellar mass of each galaxy given in Table \ref{tab:tabchar}. Appendix \ref{appendix:sedfitting} provides the details on how we processed the SED fitting by using our combined $u,g,r,i,z$ photometry plus \textit{WISE} 3.4 and 4.6 $\mu$m fluxes. The SED fitting was used to obtain only stellar masses of our galaxies, whereas the stellar population fits to the SALT spectra using full--spectrum fitting over our entire wavelength range were used to derive the galaxy ages (see Section \ref{section:analysis}). And those derived masses were then used to select our sample.  We further constrained the selection criteria by requiring that targets should show no optical emission lines by visual inspection.
 
We also included  SDSS J013403.82+004358.8 in our sample. This target was selected from the MegaZ--LRG photometric redshift catalogue \citep{Collister07}, and it has been spectroscopically confirmed as a central cluster galaxy of a very rich cluster \citep{Szabo11, Wen12}. The mass of this object was derived by fitting only its model $u, g, r, i, z$ fluxes. 

The names, coordinates and the characteristics for all of the LRGs candidates being studied in this work are summarized in Table \ref{tab:tabchar}.

\subsection{Spectroscopic Observations}

The spectroscopic observations of this sample were carried out with the 
\citep[RSS, ][]{Burgh03,Kobulnicky03} at SALT.  Long--slit optical spectra of the sample were obtained between 2011 December and 2012 September under proposal codes 2011-3-RSA\_OTH-026 and 2012-1-RSA\_OTH-013 (PI: A. Ratsimbazafy). The observation was made using the PG0900 grating at two different grating angles to cover the $\sim$4000 -- 6000 \AA{} range (rest--frame wavelength range) at both redshifts with a spectral resolution of $\sim$4--6 \AA{}.  The log of observations is given in Table \ref{tab:obslog}. A slit width of 1 arcsec, yielding a resolution of $R \sim 1900$, was used during the 2011 observations; this was increased to a slit width of 1.5 arcsec that yielded a resolution of $R \sim 1300$ during the second semester of 2012 observations.

For wavelength calibration, spectra of a Neon arc lamp were taken after each observation, as well as flat--field images in order to perform a standard reduction of two--dimensional long--slit spectra. The spectrophotometric standard stars were observed for flux calibration.

\subsection{Spectroscopic Data Reduction}

For all observations, fidelity checking and basic calibrations such as the overscan, gain, cross--talk corrections and mosaicking were already performed by the automated SALT \textsc{pyraf} pipeline called \textsc{pysalt}\footnote{\url{pysalt.salt.ac.za}} \citep{Crawford10b}. Further reduction was performed by following the standard long--slit data reduction technique with \textsc{iraf}\footnote{Image Reduction and Analysis Facility, a software system distributed by the National Optical Astronomy Observatories (NOAO). See \url{http://iraf.noao.edu/}}. Cosmic rays were removed from the two--dimensional spectrum using the \textsc{lacosmic}\footnote{An algorithm for robust cosmic ray identification using Laplacian edge detection.} (LAplacian COSMIC ray identification) software \citep{VanDokkum01}. The two--dimensional science, flat--field and arc images were trimmed to exclude nonuseful areas before performing the rest of the reduction. The two CCD gaps were filled with interpolated pixel values. All the spectra were flat--fielded by using calibrating flat--fields obtained for the science frames. This process helps to correct for the differences in sensitivity across the field and between detector pixels.  The flat--fields were also used to remove fringing at the far--red portion of the CCD chips.

The \textsc{long} package from \textsc{iraf} was then used for the wavelength calibration, extracting the spectrum, and subtracting the night sky spectrum. Spectrophotometric standard stars were reduced according to the same procedure, and a relative flux calibration was applied to the science spectrum to recover the spectral shape. Due to the nature of the SALT telescope, the unfilled entrance pupil of the telescope moves during the observation and an absolute flux calibration is not possible. We combined the individual frames to create a master science image as well as an error image. This procedure was done before extracting the one--dimensional spectra. All spectra were corrected for foreground Galactic extinction using the reddening maps of \cite{Schlegel98} and the extinction curve of the \citet{Fitzpatrick99} with $R=3.1$. The galactic $E(B-V)$ parameter for each galaxy is given in Table \ref{tab:tabchar}. 

\begin{table*}
 \begin{footnotesize}
  \caption{Results of SSP fit with BC03 models showing the SSP equivalent ages. Errors of each parameter are from the covariance matrix. S/N ratio per resolution element of the observed spectra is also given. The first nine galaxies all belong to one redshift bin that is at $z \simeq 0.40$, and the last five galaxies are in the redshift bin of $z \simeq 0.55$.}
  \begin{tabular}{lcccccl}
  \hline
  
  Name & Redshift & Age  & [Fe/H] &$\chi^2$&S/N ratio$^{\star}$\\
   
        &&(Gyr)&(dex)&&($\rm \AA^{-1}$)\\
  
 \hline

 2SLAQ J081258.12$-$000213.8&0.4063&~3.85$\pm$0.71&~0.31$\pm$0.06 &1.63&22\\
 2SLAQ J081332.20$-$004255.1&0.3924&~3.93$\pm$0.92&~0.32$\pm$0.06 &1.40&20\\
 2SLAQ J100315.23$-$001519.2&0.3980&~2.84$\pm$0.35&~0.23$\pm$0.16 &1.33&14\\
 2SLAQ J100825.72$-$002443.3&0.3984&~3.50$\pm$0.72&~0.29$\pm$0.10 &1.73&23\\
 2SLAQ J134023.93$-$003126.8&0.3997&~3.64$\pm$0.56&~0.34$\pm$0.05&1.46&33\\
 2SLAQ J134058.83$-$003633.6&0.4097&~2.82$\pm$0.11&~0.23$\pm$0.05&1.60&29\\
 2SLAQ J144110.62$-$002754.5&0.4003&~3.79$\pm$0.34&~0.26$\pm$0.03 &1.85&58\\
 2SLAQ J010427.15$+$001921.5&0.4071&~4.63$\pm$0.48&~0.12$\pm$0.02&1.63&39\\
 2SLAQ J022112.71$+$001240.3&0.3975&~3.43$\pm$0.82&--0.17$\pm$0.07&1.45&22\\
 SDSS J013403.82$+$004358.8&0.4092&~6.36$\pm$0.92&~0.24$\pm$0.02 &1.83&46\\
 2SLAQ J092612.79$+$000455.8&0.5411&~3.63$\pm$0.84&~0.14$\pm$0.10 &1.68&10\\
 2SLAQ J092740.75$+$003634.1&0.5480&~2.83$\pm$0.25&--0.33$\pm$0.07 &1.76&18\\
 2SLAQ J100121.88$+$002636.4&0.5549&~0.94$\pm$0.08&~0.29$\pm$0.06 &1.50&12\\
 2SLAQ J100131.77$-$000548.0&0.5464&~1.02$\pm$0.03&~0.07$\pm$0.08 &1.79&23\\
 2SLAQ J104118.06$+$001922.3&0.5465&~2.88$\pm$0.13&--0.11$\pm$0.09 &1.59&17\\
 2SLAQ J225540.39$-$001810.7&0.5512&~5.47$\pm$0.66&~0.07$\pm$0.04 &1.71&25\\

\hline
\multicolumn{5}{l}{$^{\star}$ S/N ratio determined at 4750 \AA{} rest--frame wavelength.}
 \label{tab:fitresults}
\end{tabular}
\end{footnotesize}
\end{table*}

\begin{figure*}
\centering
\begin{minipage}{140mm}
\epsfig{file=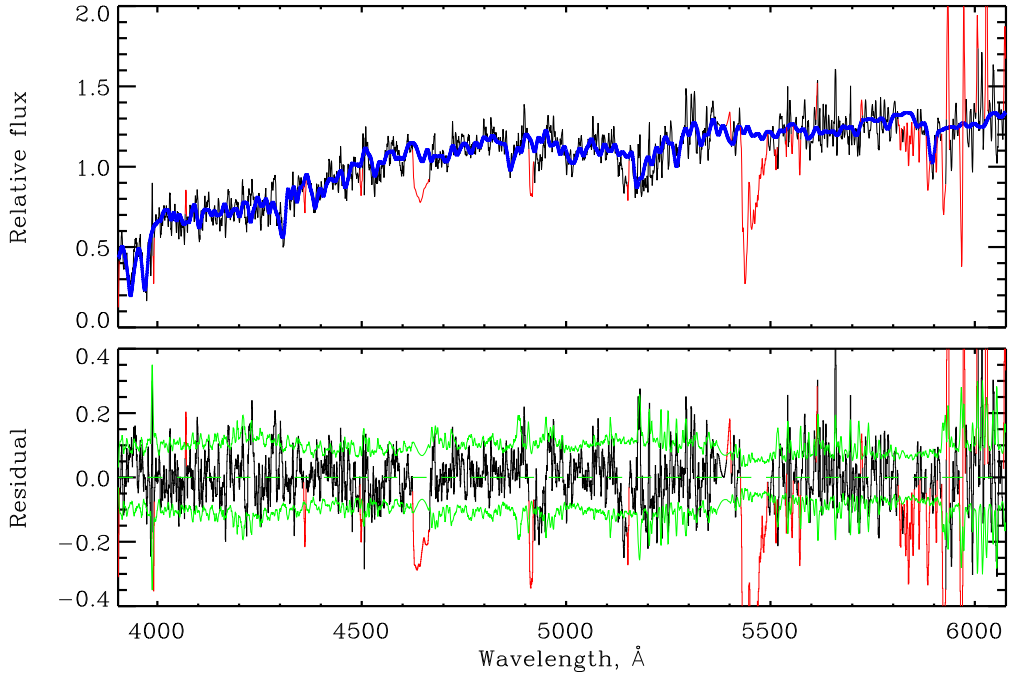,width=1.0\linewidth,clip=} 
\caption{This shows an example of full--spectrum fitting. The fitting of 2SLAQ J100825.72$-$002443.3 spectrum (smoothed with a 5 pixel boxcar) which is in black and the best fit in blue line. The red regions were excluded and masked in the fit: The outliers that correspond to the regions of the telluric lines and the residuals form the sky emission lines. The green lines in the residuals of the fit are the estimated $1-\sigma$ deviation. Fluxes are expressed in erg cm$^{-2}$ s$^{-1}$ \AA{}$^{-1}$.}
\label{plot:saltspec}
\end{minipage}
\end{figure*}

\section{Data Analysis} \label{section:analysis}

 For this work, we have used the University of Lyon Spectroscopic analysis Software \citep[\textsc{ulyss}\footnote{ \url{http://ulyss.univ-lyon1.fr}};][]{Koleva09} to measure ages for our sample of LRGs through full--spectrum fitting.  

\textsc{ulyss} compares an observed spectrum with a set of model spectra in order to derive the characteristics of the stellar population (age, metallicity and star formation history) and the internal kinematics.  It fits the entire spectrum against a model in the form of a linear combination of non--linear components (for example, age, [Fe/H] and wavelength), corrected for the kinematics and multiplied by a polynomial at the same time. The use of multiplicative polynomial makes this method insensitive to the effects of the flux calibration uncertainties and the extinction. More details of this technique of fitting are found in \citet{Koleva09}

Prior to fitting our science spectra, we needed to match the resolution between the observed spectra and models. The spectral resolution matching is described by the line spread function (LSF). In this study, we used the spectrum of a standard star HD 14802 observed with SALT during the night of 2012/10/11 along with the target SDSS J013403.82+004358.8. The atmospheric parameters ($\rm T_{eff}$, log (g) and [Fe/H]) of this star are already known in SIMBAD\footnote{\url{http://simbad.u-strasbg.fr/simbad/}} data base. Its parameters have been measured by several scientists but we chose the most recent ones by \cite{Ramirez13}. The relative LSF was obtained by comparing the standard star spectrum with the model spectrum (stellar library) with the same parameters. We used an overlapping windows of 400 \AA{} separated with 200 \AA{} steps. The decrease in instrumental velocity dispersion, $\sigma\rm_{inst}$, is typically from 110 (blue) to 70 km s$^{-1}$ (red), which is the characteristic of the spectrograph and grating, while the radial velocity, $\rm v_{rad}$ changes from 50 to 80 km s$^{-1}$, due to the uncertainty in the wavelength calibration. This relative LSF was then injected to the models to generate the resolution--matched models. Further descriptions of the model used are given below.

After matching the resolutions, we fit each LRG spectrum to single--age, single--metallicity population synthesis models.  LRGs are believed to form most of their stars very early in the Universe, and as such, single stellar population (SSP) models are generally found to be good descriptions for their spectra \citep[e.g.][]{Jimenez03,Carson10,Stern10a,Liu12,Zhang14}.  Although in our previous work \citep{Crawford10a}, we did find that LRGs may be better described by slightly extended star formation histories, we have fitted SSPs to the LRG spectra in order to compare this work to previous studies. We also fitted two stellar components to the LRG spectra in Section \ref{section:discussion}.    

The SSP models that we used are based on \textsc{galaxev} models \cite[][hereafter BC03]{BC03}, which were generated using the \textsc{stelib} library \citep{LeBorgne03} with a resolution about 3 \AA{} (full width at half--maximum) across the whole spectral range of 3200 -- 9500 \AA{}, the Chabrier initial mass function \citep[IMF;][]{Chabrier03} with a mass of 0.1 -- 100 $\rm M_{\odot}$ and have a slope of --1.35 and Padova 1994 \citep{Bertelli94} isochrones. These models cover the age from 0.1 to 20 Gyr and [Fe/H] from --2.3 to 0.4 dex, consisting of 696 different SSP spectra in total (116 number of ages and 6 metallicities). In order to be consistent with previous works on CC, the BC03 models have been chosen among the other models \citep[e.g.][]{LeBorgne04,Vazdekis10,Maraston11}. 

The red end for some of the spectra was significantly affected by fringing. Numerous and strong sky emission lines are also present in that region. Due to the presence of the fringes, removal of the sky emission  became problematic. Thus, the residuals from the bright sky emission lines are very significant. For those spectra where the residuals of the sky emission lines removal are very significant, we set the maximum rest--frame wavelength range to be 5750 \AA{}. And we did not find any change on the results while imposing the same restriction on the other spectra.

We report results of the SSP--derived ages in Table \ref{tab:fitresults}. An example of an individual fit is shown in Figure \ref{plot:saltspec}.  The errors on the parameters are the $1 \sigma$ errors. These errors are computed from the covariance matrix by the \textsc{ulyss} fitting function, which uses the \textsc{mpfit}\footnote{\url{http://cow.physics.wisc.edu/~craigm/idl/idl.html}} algorithm. In addition, \textsc{ulyss} provides the possibility of exploring and visualizing the parameter space with $\chi^{2}$ maps, convergence maps and Monte Carlo simulations to validate the errors on the parameters. We performed Monte Carlo simulations to carefully test the reliability of the fitting. A series of 500 Monte Carlo simulations was performed. Monte Carlo simulations repeated full--spectrum fitting 500 times with a random noise equivalent to the estimated noise added to the spectrum at each time. The dimension of the added noise was defined as the signal--to--noise (S/N) ratio estimates and not from the error spectra. The outputs from this simulation are the mean values of the resulting distributions of all parameters: age, metallicity, velocity dispersion and their corresponding standard deviations.   

\begin{figure}
\includegraphics[width=85mm]{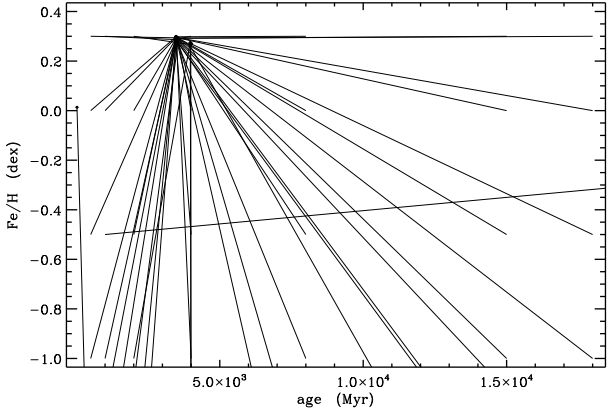}
\caption{Example of the convergence map from fitting 2SLAQ J100825.72$-$002443.3 spectrum. The results converge to the best--fitting results presented in Table \ref{tab:fitresults}.}
\label{plot:ageconv}
\end{figure}

\begin{figure}
\includegraphics[width=85mm]{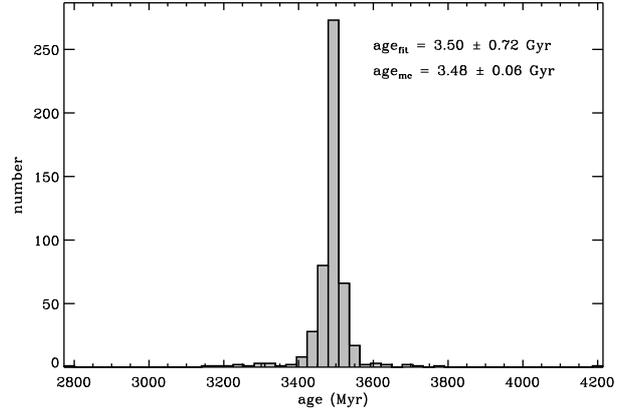}
\caption{Results from the 500 Monte Carlo simulations of 2SLAQ J100825.72$-$002443.3. This shows the histogram of the distributions of the age results from the simulations. Age values both from best fits (fit) and simulations (mc) are shown in the legend.}
\label{plot:agechi}
\end{figure}

To help assess the quality of the fits, $\chi^{2}$ and convergence maps were examined for all sources. An example of a convergence map is presented in Figure \ref{plot:ageconv}. The convergence map is presented in the age--metallicity plane. It confirms the results from the best fits and its independence from the grid of the initial guesses for age and metallicity. The plot shows the convergence regions and the location of the absolute minimum. In addition, the results of 500 Monte Carlo simulations of 2SLAQ J100825.72$-$002443.3 in the form of a histogram recovering values of age are shown in Figure \ref{plot:agechi} as an example. The Monte Carlo simulations help us to verify our SSP results by visualizing any degeneracies. The mean values of the simulations are consistent with the best--fitting results in most cases. Nonetheless, there can be some discrepancies between the two results \citep{Koleva09}, which is due to the degeneracy between parameters and the level of the noise in the spectra. In addition, the Monte Carlo simulations manage to reproduce the error spectra and give a good estimate of the errors. The errors determined by the Monte Carlo simulations are compared with those from the single fits in Figure \ref{plot:agechi} as an example and in Figure \ref{plotappendix:saltspec} in Appendix \ref{appendix:fullspetrumfitting} for all the galaxies. Table \ref{tab:fitresults} shows the estimated S/N ratios in the central extracted spectra, which are used to generate the results from the Monte Carlo simulations. These S/N ratios were estimated at a rest--frame wavelength of $\sim$4750 \AA{}.

Despite fewer number of galaxies at $z \simeq 0.55$, it is clear that there is an age--redshift relation, i.e. the mean age at $z \simeq 0.40$ is older than that at $z \simeq 0.55$. The galaxies at $z \simeq 0.40$ have an average age of $3.88 \pm 0.20$ Gyr, and $2.80 \pm 0.18$ Gyr for galaxies at $z \simeq 0.55$. The galaxies at $z \simeq 0.40$ have similar ages and metallicities, confirming that these galaxies almost form a homogeneous sample at this redshift. However, galaxies at $z \simeq 0.55$ have a large scatter in the metallicity. In this work, we find younger ages than the total sample of red galaxies in \citet{Stern10a} at both redshifts but we note that we have a small number of galaxies in our sample. The error on the mean age was obtained by applying the standard propagation of errors technique. 

The metallicity determination was not as robust as we would have liked, especially for the sample at $z \simeq 0.55$.  LRGs are found to have similar metallicities (higher than solar metallicity) and they should not evolve with redshift \citep{Jimenez03}. However, ages and metallicities at $z \simeq 0.55$ do not show consistent values in the age--metallicity space; see Table \ref{tab:fitresults}. The poorly--constrained metallicity can be the result of the quality of the spectra, low S/N ratio, or an artefact of the stellar population models using solar scale isochrones that do not predict the right abundances (in particular the enhanced $\alpha$-element abundances) for the very massive old early--type galaxies \citep{Loubser16}. Further analysis is needed to improve the results on metallicity by analysing higher S/N ratio and larger wavelength coverage spectra in which more metal--sensitive features are available. The prominent metallicity features such as Mg$_{b}$, Fe5270, Fe5335 are found to be in the extremely red part of those galaxy spectra at $z \simeq 0.55$, and where the residuals of the sky emission lines removal were significant. However, most of the absorption features (e.g. high--order Balmer lines) present in our wavelength ranges are sensitive to the age determination but insensitive to the metallicity and $\alpha$--elements (the detailed discussion about the absorption features sensitivity is given in \citet{Vazdekis15}). The advantage of using full--spectrum fitting over a number of indices is that the age--metallicity degeneracy can be broken. 

To test whether the poorly--constrained metallicity has an impact on our age determination, we run an SSP fitting with a fixed metallicity for the galaxies at $z\simeq 0.55$. The metallicity was fixed at the mean value of metallicities found for the galaxies at $z\simeq 0.40$ ([Fe/H] $= 0.20$ dex). The results on ages were still consistent with the results when the metallicity is left as a free parameter, as seen in Table \ref{tab:fitresults}. From this test, we conclude that the poorly--constrained metallicity does not have a significant influence on our derived ages.

\section{$H(\lowercase{z})$ measurement} \label{section:hz}

\begin{figure*}
\centering
\includegraphics[width=170mm]{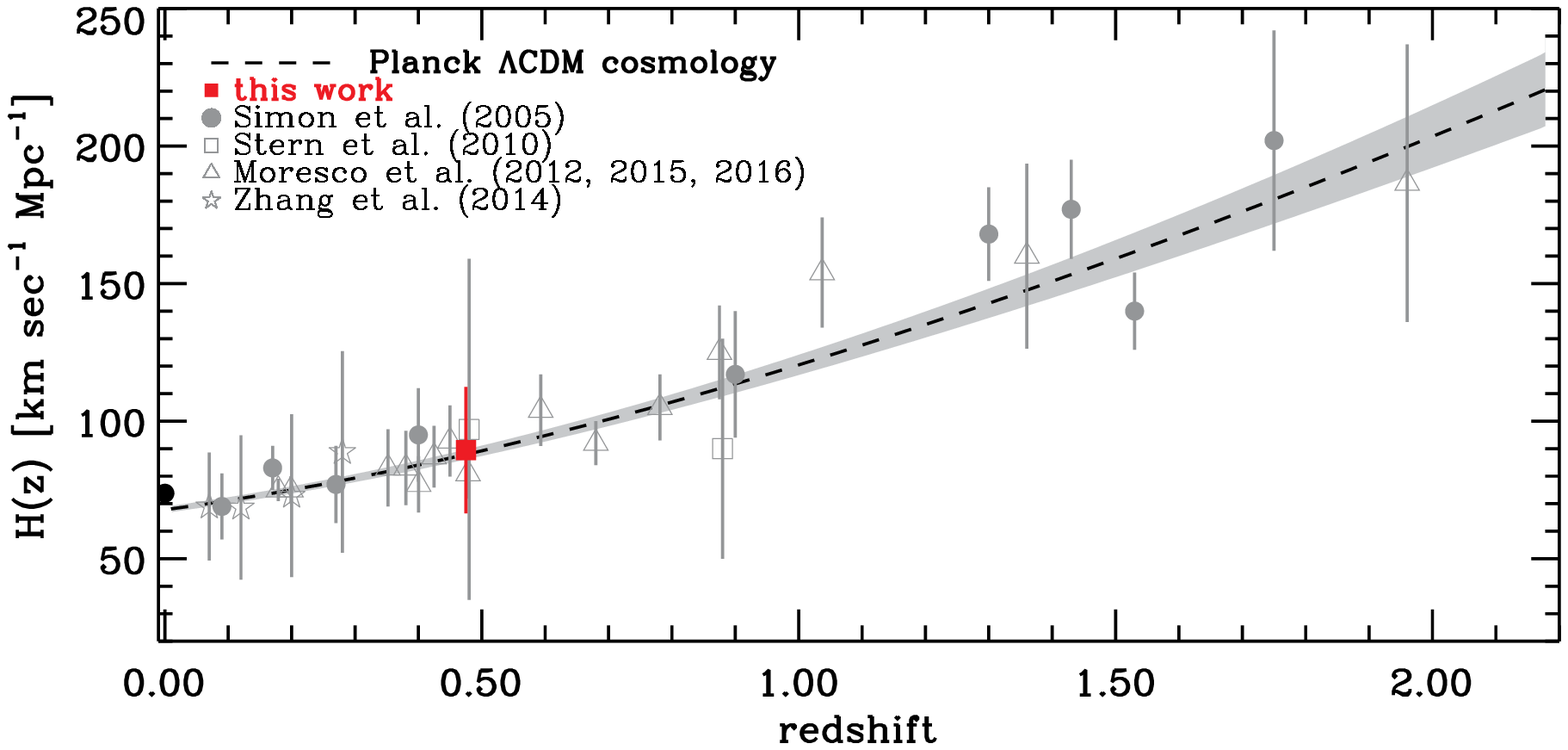}
\caption{Our estimate $H(z \simeq 0.47)$ measured using SALT LRG spectra is represented by the red filled rectangle. It has a value of $H(z) = 89 \pm 23$ (stat) km s$^{-1}$ Mpc$^{-1}$. Our result is plotted with all available $H(z)$ in the literature up to redshift $z \sim$ 2 \citep{Simon05,Stern10a,Moresco12,Zhang14,Moresco15,Moresco16}. The dashed line and the shaded regions are not a best fit to the data but the theoretical $H(z)$ of a flat $\varLambda$CDM model with its 1$\sigma$ uncertainty obtained by \citet{Planck Collaboration 2015} (with $\varOmega_{m} = 0.308 \pm 0.012$, and $H_{0} = 67.8 \pm 0.9$ km s$^{-1}$ Mpc$^{-1}$). The black point at $z=$0 is the Hubble Space Telescope measurement of the Hubble parameter today $H_{0} = 73.8 \pm 2.4$ km s$^{-1}$ Mpc$^{-1}$ \citep{Riess11}.}
\label{plot:hzval}
\end{figure*}
 
If the luminosity--weighted SSP equivalent represents the ages of the galaxies, we can use the estimated ages to calculate the Hubble parameter $H(z)$ via the difference in ages associated with the corresponded difference in redshifts ($\varDelta z/\varDelta t$).  The mean age of the sample at each redshift was used to measure the differential ages. We considered galaxies with sufficient S/N ratio which are all shown here (minimum S/N ratio is 10 \AA$^{-1}$). 

Applying the Equation \ref{hzeqn} for $z\simeq 0.47$, the redshift between $z \simeq 0.40$ and $0.55$, we obtained a new observational Hubble parameter $H(z\simeq 0.47) =89 \pm 23$ (stat) km s$^{-1}$ Mpc$^{-1}$. We plot this latter with all available observational $H(z)$ values \citep{Simon05,Stern10a,Moresco12,Zhang14,Moresco15,Moresco16} up to $z \sim 2$ derived using the CC methodology in Figure \ref{plot:hzval}. The most comparable measurements at the same redshift are by \citet{Stern10a}, who measured a value of $H(z) = 97 \pm 60$ km s$^{-1}$ Mpc$^{-1}$ at $z=0.48$, and by \citet{Moresco16} who derived a measurement of $H(z) = 80.9 \pm 9$ km s$^{-1}$ Mpc$^{-1}$ at $z=0.478$. Our value is consistent with the standard cosmology model with parametrization of $\varOmega_{m} = 0.308$, and $H_{0} = 67.8$ km s$^{-1}$ Mpc$^{-1}$ by \citet{Planck Collaboration 2015}.  

There are two main sources of errors when determining the Hubble parameter $H(z)$. The statistical error that is related to the age measurement itself. It is related to the covariance matrix for the full--spectrum fitting. The systematic error should not be also 	underestimated. A description of possible sources of the systematic errors is given below, however, the details of their determination are given in Section \ref{section:discussion}. 

Statistical error: The error in $H(z)$ depends only on the age at each redshift since the error on the redshift is negligible, and was calculated from 
\begin{equation}
\dfrac{\sigma^{2}_{H}}{H(z)^{2}} = \dfrac{(\sigma^{2}_{t_{1}} +\sigma^{2}_{t_{2}})}{(t_{1} - t_{2})^2}
\end{equation}
where $t$ is the age at each redshift and $\sigma_{t}$ is the error on that age. 

Systematic error: There are many sources of systematic errors as discussed by various authors who have worked on CC technique \citep[e.g.][]{Jimenez03,Moresco12}. Here we consider the effect of the stellar population model adopted and the star formation history contribution. First, to consider the uncertainty due to the different SSP models, we fitted our spectra with four SSP models and estimated the values of $H(z)$. The systematic uncertainty on $H(z)$ would be the standard deviation between these $H(z)$ estimates. Secondly, despite the fact that all galaxies studied here are classified as passive, their star formation histories may show different bursts including a fraction of younger stellar populations. The systematic uncertainty due to the contribution of possible young stellar populations would be the mean of the light fraction of these populations. The total systematic error on $H(z)$ would be determined by adding the two uncertainties in quadrature .

\section{Discussion} \label{section:discussion}
\subsection{Sensitivity to the stellar population model}
In this paper, we use the \textsc{galaxev}/\textsc{stelib} models by \citet{BC03} to fit the observed spectra. These models are widely used by authors working on CC. We also tested other models such as Pegase--HR/ELODIE 3.1 \citep[][PE]{LeBorgne04}, Vazdekis/MILES \citep[][VM]{Vazdekis10} and the latest Maraston/MILES models \citep[][M11]{Maraston11}, in order to check any systematic differences. Each model has its own ingredients and settings to build the parameter grids (IMF, stellar library, wavelength range, sampling and coverage of stellar parameter space, etc.). The metallicity coverage of models using MILES library is limited to lower metallicity than those models using different libraries. When using models based on the MILES library, some of the LRGs are found to have higher metallicity (see Table \ref{tab:fitresults}), and some galaxies thus hit the upper limits of the metallicity values. Moreover, M11 models have a poor sampling of stellar parameter space at the lowest metallicity. We were forced to further reduce the range of parameters of these models in order to conform with \textsc{ulyss} format. At the end, the M11 models cover  the age from 0.1 to 15 Gyr and metallicity from --1.30 to 0.30 dex (25 number of ages and 4 metallicities).

\begin{figure}
\centering
\includegraphics[width=85mm]{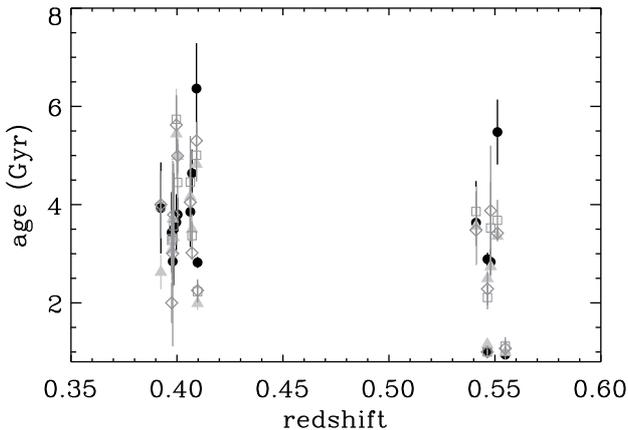}
\caption{Different SSP equivalent ages derived from \textsc{galaxev}/\textsc{stelib} models (black filled circles), Pegase--HR/ELODIE (grey filled triangles), Vazdekis/MILES models (grey open squares) and Maraston/MILES (grey open diamonds) at each redshift.}
\label{plot:allges}
\end{figure}

We found that there is some model dependence. All four models are probably different among each other over the whole wavelength range. Therefore there are always some discrepancies in the parameter outputs of these models. Figure \ref{plot:allges} illustrates the different SSP equivalent ages using \textsc{galaxev}/\textsc{stelib}, PE, VM and M11 models. All four models can give more or less reasonable results and they are comparable to each other. The model dependence obviously provides some changes on $H(z)$ estimates. The dispersion between the $H(z)$ measurements can then quantify the systematic errors. Despite using different ingredients, the systematic errors on $H(z)$ among the four models were at $\sim$8\% level. Because we used models with fairly similar ingredients, our estimation of the systematic error should be regarded as a lower limit. If models with very different ingredients were used, then the systematic would most likely be larger. 

\subsection{Reconstruction of the star formation histories}

According to the selection criteria, all LRGs should show passively evolving stellar populations, i.e. their stellar populations are best represented by SSP equivalent ages and metallicities. The derived ages and metallicity are light--weighted since the SSP represents the light--dominant epoch of star formation. The young ages obtained for some galaxies may be due to the presence of a small amount of residual star formation. In these cases, an SSP is not an entire accurate approximation and the presence of a small young stellar population will lead to younger SSP--equivalent ages. 

\textsc{ulyss} does not provide the exact star formation rate (SFR) after fitting; however, it gives the optimal weights of the SSP components that we can convert as a smoothed SFR following the approach adopted by \citet{Koleva09b} and \citet{Groenewald14}. We analysed our sample with a model of two components, which is defined by the age limits of [10,1000], [1000,10000] Myr, and a free metallicity. The aim of this exercise is to reconstruct the scenarios of the star formation, to detect if there is a young population within the observational errors and to compare with the results from SED fitting. 

Considering all of the galaxies, the SFR was constructed using BC03 models. We also tested it with the others models; the results of fits are consistent using all sets of models. Most of our galaxies are very well approximated by an SSP. Those galaxies have negligible light fractions and the age of the young population $t\rm_{young}$ less than $\sim$0.1 Gyr. And their young population component does not contribute to the total stellar mass whatsoever because the mass fraction is less than $\sim$1\%. However, in 2SLAQ J100121.88$+$002636.4, a young component was found with the light fraction greater than 20\%. 2SLAQ J100121.88$+$002636.4 has a young population around $t\rm_{young}=$ 0.81 Gyr in combination with a slightly older population of $t\rm_{old}=$ 1.77 Gyr which is still young compared to the others. These results are displayed in Table \ref{tab:reliabilityresults}. This detectable burst that occurs within 1 Gyr contributes 16\% to the total stellar mass. This scenario illustrates the fact that there is a continuous star formation in this galaxy. Regardless of its classification, the stellar population of this galaxy suggests that it might be represented by those LRGs with a continuous SF \citep{Roseboom06}. It is important to note that this galaxy is the faintest one in $g$--band, and such faint galaxies are known to slightly deviate from pure passive evolution \citep{Tojeiro10}.  

When we fitted  2SLAQ J100131.77$-$000548.0  with the two burst models by imposing the young population to be younger than 1 Gyr, the fit was limited at the 1 Gyr--age limit that indicates that the other box dominates the light. We therefore changed the limits of the young component to be [10,2000] Myr. We found that there is a young population of $t\rm_{young}$ 1.13 Gyr, which dominates the light at 69 \% level and contributes to the stellar mass at 63\% level, and an old population of $t\rm_{old}$ 2.98 Gyr which contributes about 37\% to the rest of mass and  about 31\% to the rest of the light (see Table \ref{tab:reliabilityresults}). This can explain the SSP equivalent age of 1.02$\pm$0.03 Gyr as the light--dominant epoch of star formation.  

On the other hand, these two galaxies do not show any recent star formation in the SED fitting since we have assumed a single exponentially decreasing SFR as an old stellar population, i.e. no recent burst (see Appendix \ref{appendix:sedfitting} for more details). But even when considering both old and young populations, the recent SFR parameters (the stellar mass fraction due to the young stellar population, frac$\rm_{burst}$; and the mass--weighted age of the young stellar population, age$\rm_{ySP}$) with \textsc{cigale} are not very well constrained, for example in \citet{Malek14}. Given that and the poor fit on using a combination of the two bursts, the contribution of the young stellar population to the stellar mass, frac$\rm_{burst}$, is typically not greater than 24\% for these two galaxies and the mass--weighted age of young stellar population age$\rm_{ySP}$ is less than 0.3 Gyr. 

We can confirm that the method of the full--spectrum fitting is very reliable in detecting residual star formation in the galaxy compared to the method of the SED fitting we have used here. Similar to the SSP fitting discussed in Section \ref{section:analysis}, we also checked the reliability of the fits. Here we discussed the special cases of 2SLAQ J100121.88$+$002636.4 and 2SLAQ J100131.77$-$000548.0 in which non--negligible fractions of the young component are found. We performed 500 Monte Carlo simulations for both galaxies to check the stability of the fits. The Monte Carlo estimated values of the parameters are consistent with those of the single fit within the errors bars for both components, and this gives some confidence in the results.  For 2SLAQ J100121.88$+$002636.4 and 2SLAQ J100131.77$-$000548.0, the estimated values of the young and old ages are $t\rm_{young}=$ 0.70$\pm$0.35 Gyr, $t\rm_{old}=$ 1.81$\pm$1.06 Gyr and $t\rm_{young}=$ 0.98$\pm$0.01 Gyr, $t\rm_{old}=$ 3.23$\pm$0.93 Gyr, respectively. We further tested the robustness of the solutions by fitting with different models. Table \ref{tab:reliabilityresults} illustrates the luminosity--weighted age results of the two components using different models. There is a satisfactory agreement in term of age values, despite the different ingredients of the models. Although, the light fraction of each component is slightly different for each model. 

In short, we provide here the systematic error that the presence of young stellar populations is potentially contributing to the estimation of $H(z)$. When all of the LRGs are included, then the systematic error of 49\% may be introduced to the $H(z)$ estimation. If we remove the two young galaxies at $z\simeq$0.55, $H(z\simeq 0.47)$ has a very large value of $554\pm1101$ km s$^{-1}$ Mpc$^{-1}$. This is because the sample at $z\simeq$0.55 is too small and it is not statistically reliable. One of the results of this pilot study is the emphasis of larger samples at $z\simeq$0.55 (see the Conclusions).

 \begin{table}
  \caption{Luminosity--weighted ages from two populations (young YSP and old OSP) fit using four different models. The case study of 2SLAQ J100121.88$+$002636.4 and 2SLAQ J100131.77$-$000548.0. The light fraction for each component is also shown in bold. Ages are given in Myr.}
  \begin{tabular}{ccc|cc}
  \hline
\multicolumn{1}{c}{}&
\multicolumn{2}{c}{2SLAQ J100121.88$+$002636.4} &
\multicolumn{2}{c}{2SLAQ J100131.77$-$000548.0}\\
  
  Model          &   YSP                             & OSP                                     &     YSP                   &                    OSP  \\
   \hline
  		     &    \textbf{30\%} 		    &   \textbf{70\%}                        &  \textbf{69\%} 		    &   \textbf{31\%} \\
  BC03           &	   	817$\pm$491	    &     1771$\pm$267                  &	1139$\pm$83	            &	2984$\pm$818		\\
  		     &    \textbf{31\%} 		    &   \textbf{69\%}                        &  \textbf{43\%} 		    &   \textbf{57\%} \\
  PE               &		43$\pm$30           &     1001$\pm$214                   &	1179$\pm$167		    &	3949$\pm$635		\\
  		     &    \textbf{0\%} 		    &   \textbf{100\%}                       &  \textbf{77\%} 		    &   \textbf{23\%} \\
  VM              &					    &     1122$\pm$158                   & 999$\pm$104	            &	2143$\pm$809		\\
  		     &    \textbf{20\%} 		    &   \textbf{80\%}                        &  \textbf{92\%} 		    &   \textbf{8\%} \\
  M11             &	  703$\pm$354		    &    1242$\pm$302                    & 1051$\pm$84	            &	3454$\pm$948		  \\
 
   \hline
    \label{tab:reliabilityresults}
  \end{tabular}
  \end{table}

\subsection{Formation epoch} 

From the discussion above, we have proved that our selected galaxies are passively evolving except two that are less passive. We used the mean ages obtained above to calculate the mean age formations of these galaxies by assuming a \textit{Planck} $\varLambda$ cold dark matter cosmology with $H_{0} = 67.8 \pm 0.9$ km s$^{-1}$ Mpc$^{-1}$ and  $\varOmega_{m} = 0.308 \pm 0.012$. The mean age formation at redshift $z$ is the difference between the age of the Universe and the mean age of LRGs at the same redshift. When all LRGs are included, a mean formation age of $\sim$5.5 Gyr was obtained for both redshifts, which is very consistent with the mean formation age found in \citet{Liu12} using the technique of full--spectrum fitting when age--dating LRGs. However, \citet{Carson10} obtained a formation age of 4.5 Gyr when using another method of age--dating that is based on the use of Lick--indices. The 1 Gyr difference might be due to the different stellar populations used. Both studies extracted the SSP ages of LRGs by co--adding their spectra in each redshift bin. They stopped the age--redshift relation determination at $z \sim$0.40 of four different sub--samples according to their velocity dispersions. SSP equivalent ages obtained at the last redshift bin for both studies are comparable to our SSP--equivalent ages at $z \simeq$ 0.40. This implies that if an interpolation is drawn from those age--redshift relations, our SSP--equivalent ages at $z \simeq$ 0.55 would still be at the right range. Furthermore, we obtained average ages and masses consistent with the average ages and masses of passive LRGs obtained in \citet{Maraston13}; see their Figures 5 and 6. \citet{Barber07} estimated the light--weight age, mass, metallicity and SFH of SDSS--LRGs. They found that majority of stars in LRGs formed at redshift $z \simeq$1.1--1.9, with 80\% of their stellar masses already been assembled around $z \simeq$ 0.7--1.1. We obtained an average age formation of $\sim$5.5 Gyr, which corresponds to a formation redshift of $z \simeq$1.11. And yet this value falls into the formation redshift range found by \cite{Barber07}. The mean formation age at $z\simeq $0.55 has changed from 5.5 to 4.6 Gyr (formation redshift of 1.37) when we excluded the two young galaxies.

\section{Conclusions}

We have reduced and analysed 16 long--slit spectra of LRGs recently observed with SALT. These galaxies were selected from the 2SLAQ and MegaZ--LRGs catalogues at redshift $z \simeq 0.40$ and $0.55$. Our selection is based on stellar mass, brightness and the lack (or absence) emission lines of the galaxy in order to have a sample of old and massive passively--evolving galaxies. In this paper, we derive their ages by applying the full spectral fitting method using the \textsc{ulyss} package developed by \citet{Koleva09}. The \textsc{galaxev} models with the \textsc{stelib} stellar library are used to extract the SSP parameters.

The mean age at each redshift bin is used to measure the Hubble parameter $H(z)$ at $z \simeq 0.47$  by adopting the method of CC. We find an improved $H(z)$ measurement over the \citet{Stern10a}.

As described by \citet{Crawford10a} and also evident in studies at lower redshift, further improvements in the value of $H(z)$ can be made by increasing the sample size. The relatively small sample presented here was part of an initial pilot study and a test of this method using observed LRGs with SALT. It will contribute to complete the long--term goal of our future survey. We aim to observe more galaxies in order to improve our $H(z)$ measurement and build the evolution of the Hubble parameter as a function of redshift over $0.1 < z < 1$. Further, the current estimate $H(z)$ at $z \simeq 0.47$ is comparable to the existed value in literature, which is reassuring for the future surveys. This value will be combined with the available $H(z)$ in the literature in order to constrain cosmological parameters. 

\section*{Acknowledgments}

We thank the referee for the useful comments that helped to clarify the paper. ALR acknowledges the South African Square Kilometer Array project and the National Research Foundation (South Africa)
for financial support during this project. ALR is grateful to the Royal Society/National Research Foundation (RS/NRF) project for supporting her visit to the Institute of Cosmology and Gravitation
in Portsmouth - United Kingdom. We thank Tim Higgs, Michelle Cluver and Thomas Jarrett for some useful discussions about the catalogue matching. SMC and PV acknowledge support from the National
Research Foundation of South Africa. All the observations reported in this paper were obtained with the SALT under programmes 2011-3-RSA_OTH-026 and 2012-1-RSA_OTH-013 (PI: Ando Ratsimbazafy).
This research has made use of the SIMBAD data base, operated at CDS, Strasbourg, France, and the NASA/IPAC Extragalactic Data base (NED) which is operated by the Jet Propulsion Laboratory, California Institute of Technology, under contract with the National Aeronautics and Space Administration.



\appendix
\section{SED fitting} \label{appendix:sedfitting}
\subsection{Matching 2dF--SDSS with \textit{WISE} catalogue}
Optical and mid--infrared surveys are frequently cross--matched in order to conduct multi-wavelength studies of any extragalactic sources. For instance \citet{Donoso12} and \citet{Yan13} combined \textit{WISE} and SDSS to investigate the properties of any types of galaxies in the mid-IR regime. We followed the method described in \citet{Yan13} for the matching between 2SLAQ--LRGs and \textit{WISE}, including the radius matching of 3 arcsec and to a depth of $r$ = 22.6.  We used the SDSS model photometry and the \textit{WISE} profile-fit photometric value for further analysis. This reduces the influence of matched apertures, differences between the resolution, seeing (or point spread function, more generally), aperture size, etc., of the catalogues.

\begin{table*}
 \begin{footnotesize}
  \caption{Results of SED fitting with \textsc{cigale} for all galaxies.}
  \begin{tabular}{lcccccl}
  \hline
  
  Name & Redshift & log$M_{*}$  & log SFR & $\rm age_{M}$& D4000&$\chi^2$\\
   
        &         &($\rm M_{\sun}$)&($\rm M_{\sun} yr^{-1}$)& (Gyr)&\\
  
 \hline

 2SLAQ J081258.12$-$000213.8&0.4063&11.60$\pm$0.05&1.52$\pm$0.07&4.83$\pm$0.19&1.37$\pm$0.02&2.48\\
 2SLAQ J081332.20$-$004255.1&0.3924&11.62$\pm$0.11&2.03$\pm$1.02&4.55$\pm$0.20&1.34$\pm$0.27&0.09\\
 2SLAQ J100315.23$-$001519.2&0.3980&11.29$\pm$0.06&1.21$\pm$0.07&4.83$\pm$0.19&1.37$\pm$0.02&1.70\\
 2SLAQ J100825.72$-$002443.3&0.3984&11.41$\pm$0.06&1.34$\pm$0.08&4.80$\pm$0.20&1.37$\pm$0.02&2.94\\
 2SLAQ J134023.93$-$003126.8&0.3997&11.46$\pm$0.05&1.38$\pm$0.08&4.81$\pm$0.20&1.37$\pm$0.02&4.44\\
 2SLAQ J134058.83$-$003633.6&0.4097&11.38$\pm$0.06&1.31$\pm$0.08&4.80$\pm$0.20&1.37$\pm$0.02&1.12\\
 2SLAQ J144110.62$-$002754.5&0.4003&11.76$\pm$0.05&1.68$\pm$0.07&4.83$\pm$0.19&1.37$\pm$0.02&5.27\\
 2SLAQ J010427.15$+$001921.5&0.4071&11.52$\pm$0.06&1.45$\pm$0.07&4.81$\pm$0.20&1.37$\pm$0.02&3.71\\
 2SLAQ J022112.71$+$001240.3&0.3975&11.15$\pm$0.06&1.08$\pm$0.08&4.80$\pm$0.20&1.37$\pm$0.02&1.59\\
 SDSS J013403.82$+$004358.8&0.4092&11.96$\pm$0.13&1.55$\pm$0.44&6.39$\pm$1.13&1.54$\pm$0.20&0.08\\
 2SLAQ J092612.79$+$000455.8&0.5411&11.64$\pm$0.11&2.05$\pm$1.00&4.75$\pm$0.20&1.33$\pm$0.26&0.51\\
 2SLAQ J092740.75$+$003634.1&0.5480&11.40$\pm$0.06&1.32$\pm$0.08&4.81$\pm$0.20&1.37$\pm$0.02&5.32\\
 2SLAQ J100121.88$+$002636.4&0.5549&11.25$\pm$0.08&1.19$\pm$0.10&4.79$\pm$0.20&1.37$\pm$0.02&3.52\\
 2SLAQ J100131.77$-$000548.0&0.5464&11.51$\pm$0.06&1.44$\pm$0.08&4.81$\pm$0.20&1.37$\pm$0.02&4.80\\
 2SLAQ J104118.06$+$001922.3&0.5465&11.52$\pm$0.06&1.45$\pm$0.09&4.80$\pm$0.20&1.37$\pm$0.02&3.39\\
 2SLAQ J225540.39$-$001810.7&0.5512&11.69$\pm$0.06&1.62$\pm$0.08&4.80$\pm$0.20&1.37$\pm$0.02&3.83\\
 
\hline
 \label{tabappendix:sedfitresults}
\end{tabular}
\end{footnotesize}
\end{table*}
 
\subsection{Stellar masses}
The \textsc{cigale} code calculates a grid of theoretical SEDs and fits observed photometric fluxes from UV to IR. The choice of the input parameters is critical, depending on the aim of the study. Here we present the basic input parameters that we have used. Models are generated with a stellar population synthesis code based on \citet{Maraston05}, which is one of the two models provided, and by adopting input parameters of star formation, UV--optical attenuation and IR emission. We assumed a solar metallicity and a Salpeter IMF. It is important to note that the metallicity is fixed in this code. The star formation history implemented in \textsc{cigale} is a combination of two bursts representing an old more passively evolving and young population. We assumed an exponentially decreasing SFR over 8 Gyr, adopting an e--folding time ranging from 1 to 10 Gyr with 2 Gyr steps for the old populations. However, no recent burst was considered for young stellar populations. After applying different scenarios of SFR using both old and young populations, only the parametrization of a single old burst gave the best fits to our data. We decided not to modify the slope of the Calzetti attenuation curve and not to add any UV bump. We considered the effect of attenuation for old stellar population by adding the reduction factor $f\rm_{att}$ = 0.5. The semi-empirical model templates of \citet{Dale02} were chosen to fit IR observations that are parametrized by the power--law slope $\alpha$ in the interval [1; 2.5]. 

\begin{figure*}
\centering
\begin{minipage}{140mm}
\epsfig{file=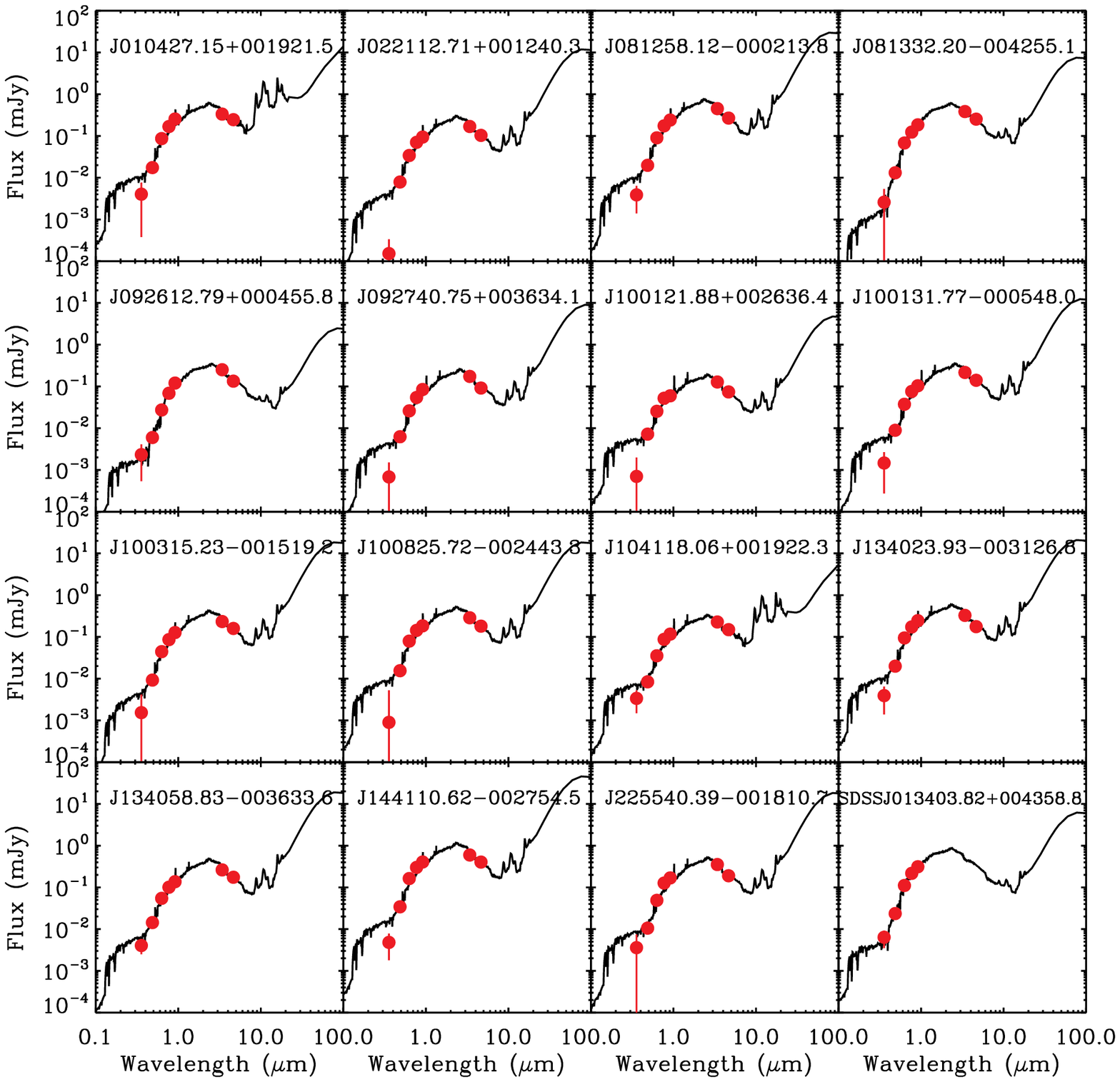,width=1.0\linewidth,clip=} 
	\caption{The SED fitting with \textsc{cigale} for all galaxies. The best--fitting models are plotted with the observed fluxes in red points, and the best--fitting model with a solid line.}
\label{plotappendix:sedfit}
\end{minipage}
\end{figure*}

Figure \ref{plotappendix:sedfit} represents the SED fitting of the 16 galaxies. Table \ref{tabappendix:sedfitresults} illustrates the output parameters (stellar mass $M_{*}$, SFR, mass--weighted age, $\rm age_{M}$, D4000 index and $\chi^{2}$) from the SED fitting. The best--fitting models superposed on the observed fluxes are shown in Figure \ref{plotappendix:sedfit}. The reduced $\chi^2$ of the fit is between 0.08 and 5.3. We can clearly see that some observed fluxes are not fitted well. For instance, the observed $u$--band flux for most of the SED fitting is not well reproduced by the code, which might be due to the fact that $u$--band flux for some galaxies is often swamped by the photon noise of the sky \citep{Blanton03}. This can be identified by the size of errors on the flux and the value of the flux itself (very faint) for some galaxies. Note that the accuracy of the output relies on the input parameters. For our sample selection, we only require an estimate of the stellar mass, therefore we did not carry out exhaustive testing of input parameters. 

Apart from the stellar mass, the derived parameters such as mass--weighted age and light--weighted age might be useful for us to compare with the SSP parameters from the full--spectrum fitting. Unfortunately, in this current version, the light--weighted age was not provided but the dust--free D4000 break  measured on the synthesized spectra was produced instead. The most appropriate comparison would be the light--weighted age since it may provide information on the age of the stellar population. The D4000 break must be large for old galaxies and small for the young populations. \citet{Kauffmann03} found that old elliptical galaxies have a typical D4000 index around 1.85 and D4000 index of $\sim$1.3 describes young populations. Our galaxies have low values of D4000 index of around 1.37 even for these massive galaxies that are different from the SDSS galaxies in \citet{Kauffmann03}, but quite similar to that of \citet{Buat11} by using the same code and models. These results are incomparable to the values found using the full--spectrum fitting. We cannot rely on these values since they are derived from models with fixed metallicity. In general the D4000 break is very sensitive to metallicity and it should not to be underestimated. 

\section{Full spectrum--fitting of all spectra} \label{appendix:fullspetrumfitting}
In this appendix, we present the full--spectrum fitting of all galaxies performed with \textsc{ulyss}. The individual fit is shown in Figure \ref{plotappendix:saltspec} as well as the results of the Monte Carlo simulations. The upper age limit for the Monte Carlo simulations was set to be the age of the Universe at each redshift. The mean values of the simulations are compared with the best--fitting results, and they mostly agree. However, there are some galaxies that could be considered exceptions to these Monte Carlo simulation results, for example, SDSS J013403.82+004358.8, 2SLAQ J134023.93$-$003126.8 and 2SLAQ J092612.79+0043.58.8. These galaxies gave different Monte Carlo results compared to the best--fitting results. SDSS J013403.82+004358.8 and 2SLAQ J134023.93$-$003126.8 also show two different possible solutions in the $\chi^{2}$ map and convergence map. Nevertheless, we did not find any evidence for the ongoing star formation in these galaxies. For 2SLAQ J092612.79+0043.58.8, it has the lowest S/N ratio and the fit failed to reproduce a well--defined error distribution. 

\begin{figure*}
\centering
\begin{tabular}{cc}
\epsfig{file=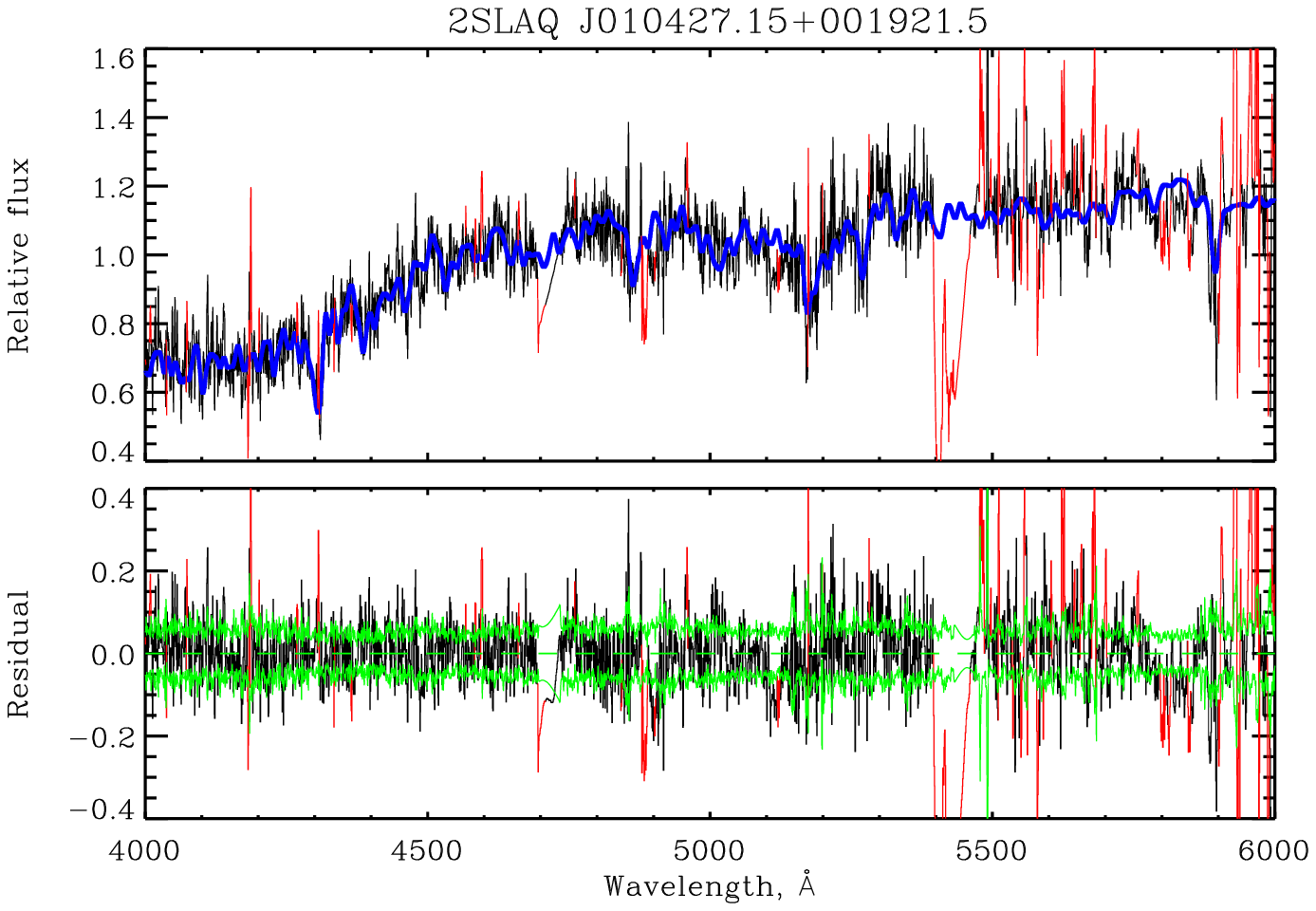,width=0.4\linewidth,clip=} & 
\epsfig{file=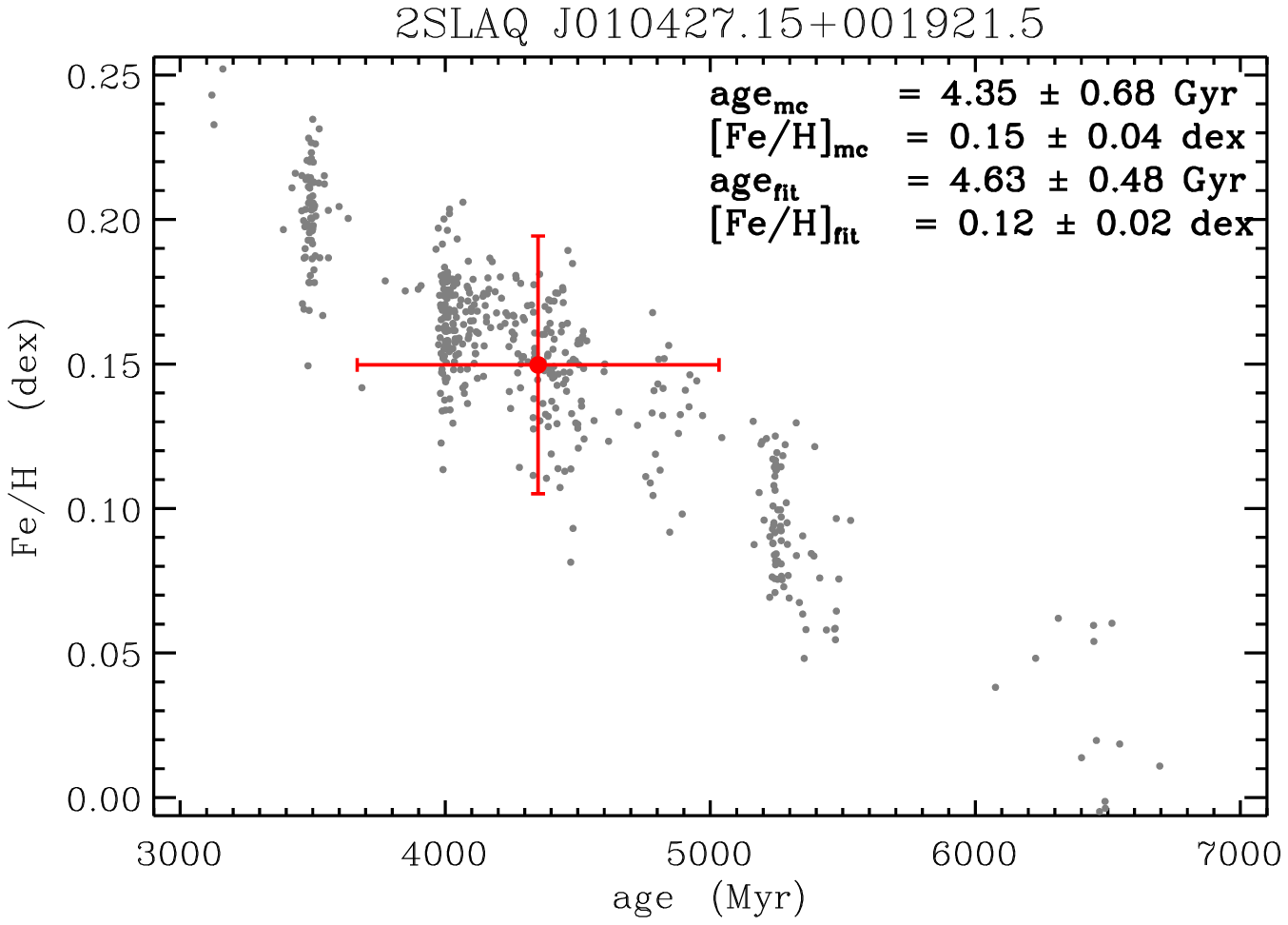,width=0.4\linewidth,clip=} \\
\epsfig{file=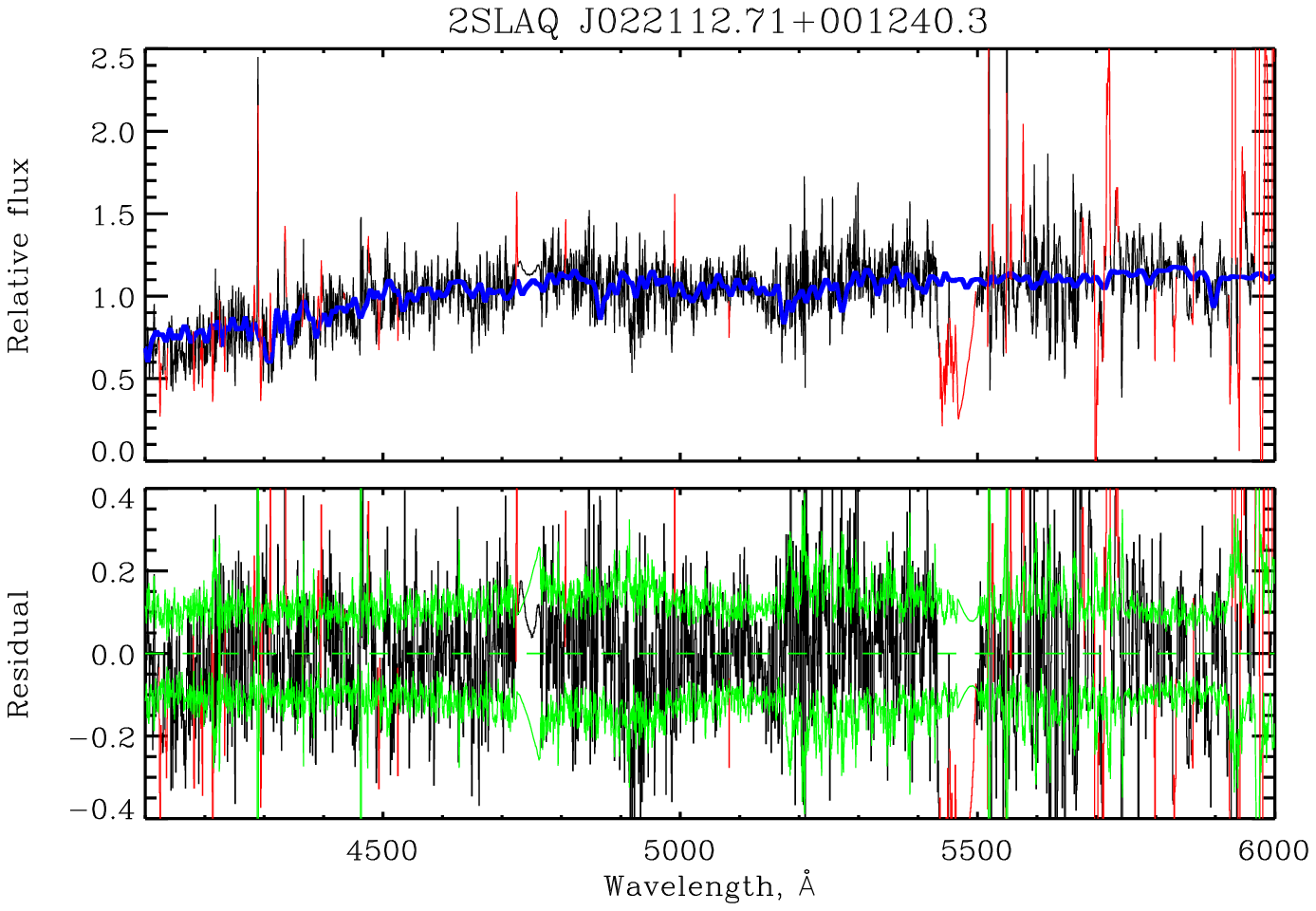,width=0.4\linewidth,clip=} &
\epsfig{file=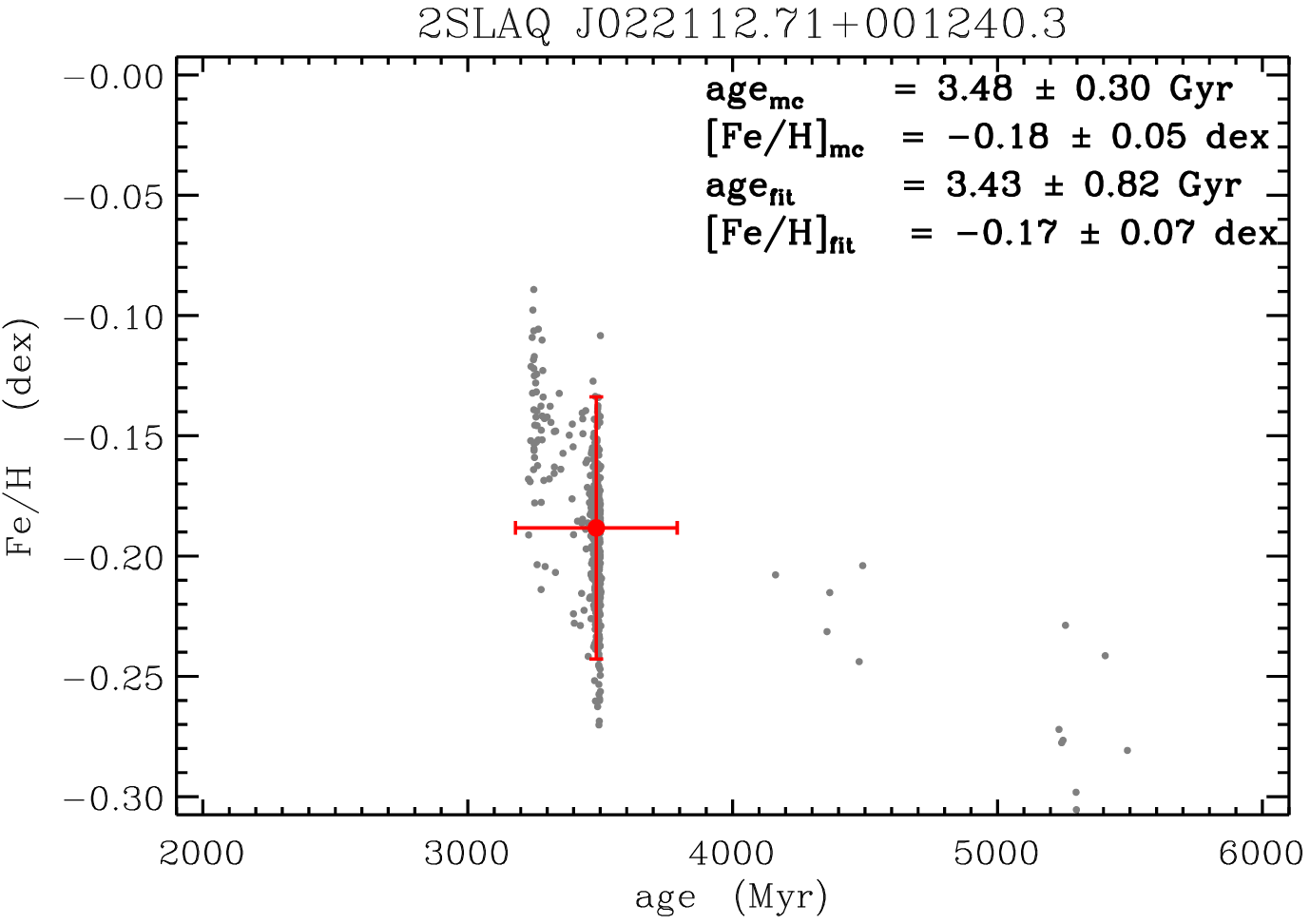,width=0.4\linewidth,clip=}\\
\epsfig{file=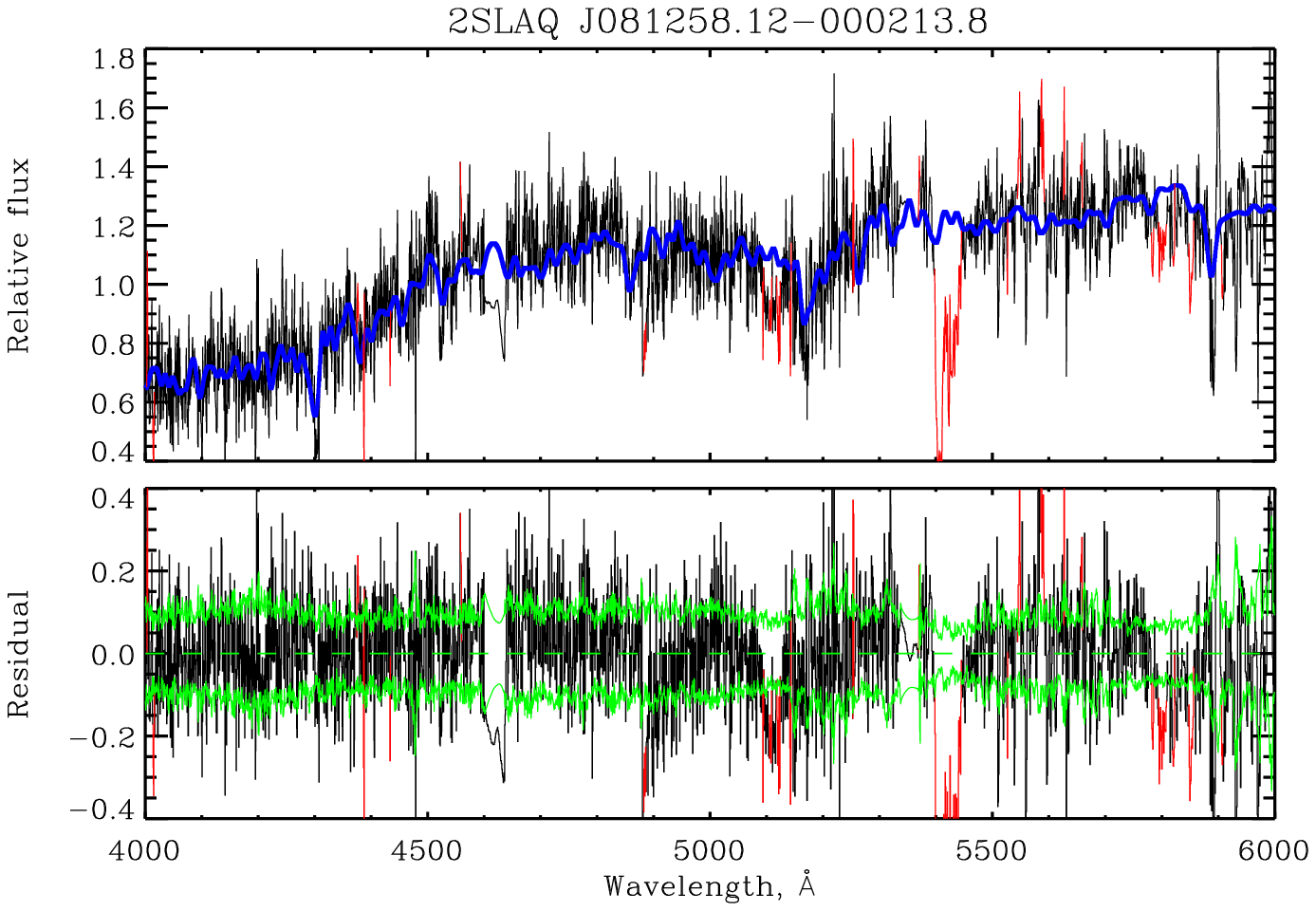,width=0.4\linewidth,clip=} &
\epsfig{file=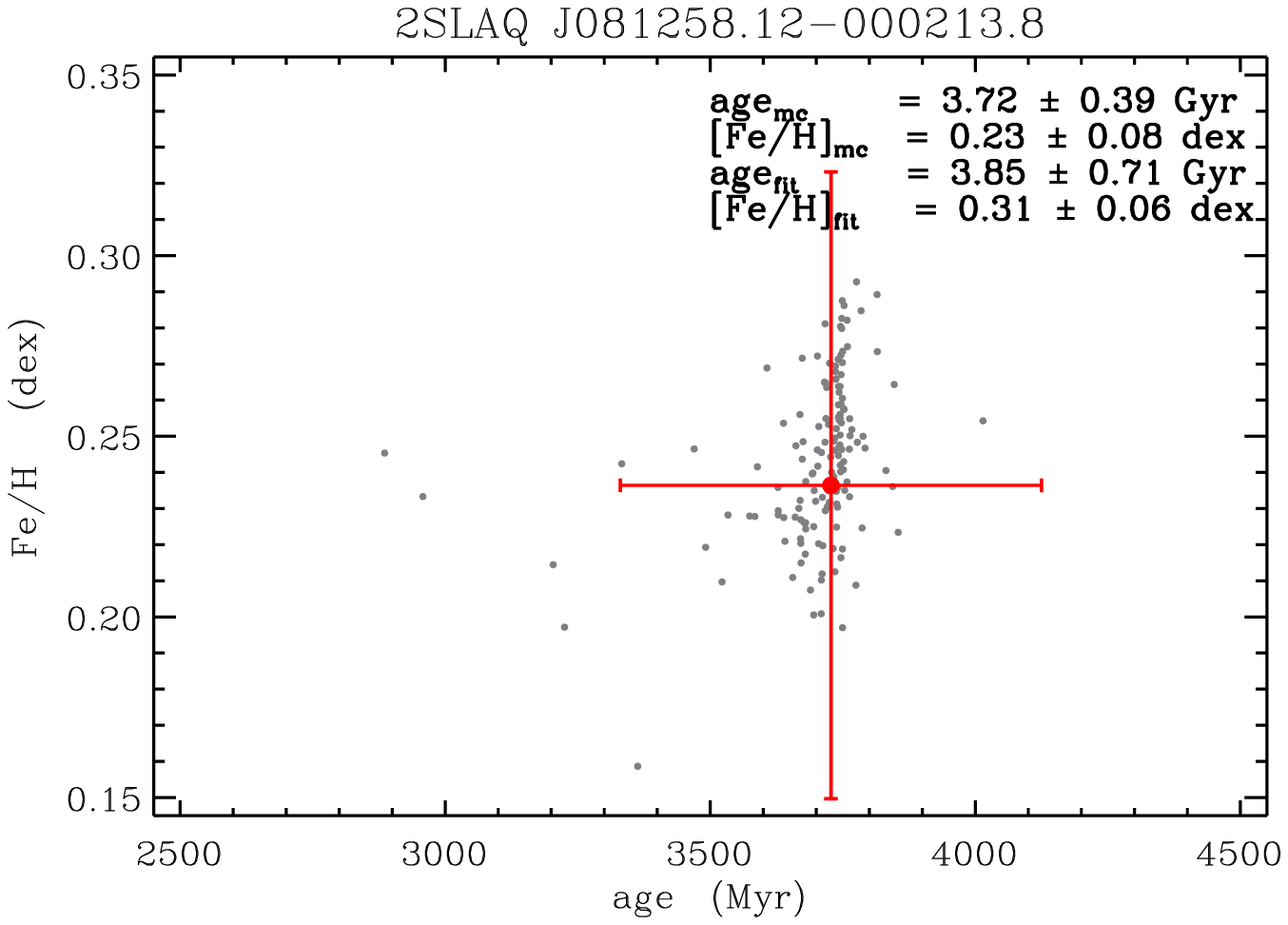,width=0.4\linewidth,clip=}\\
\epsfig{file=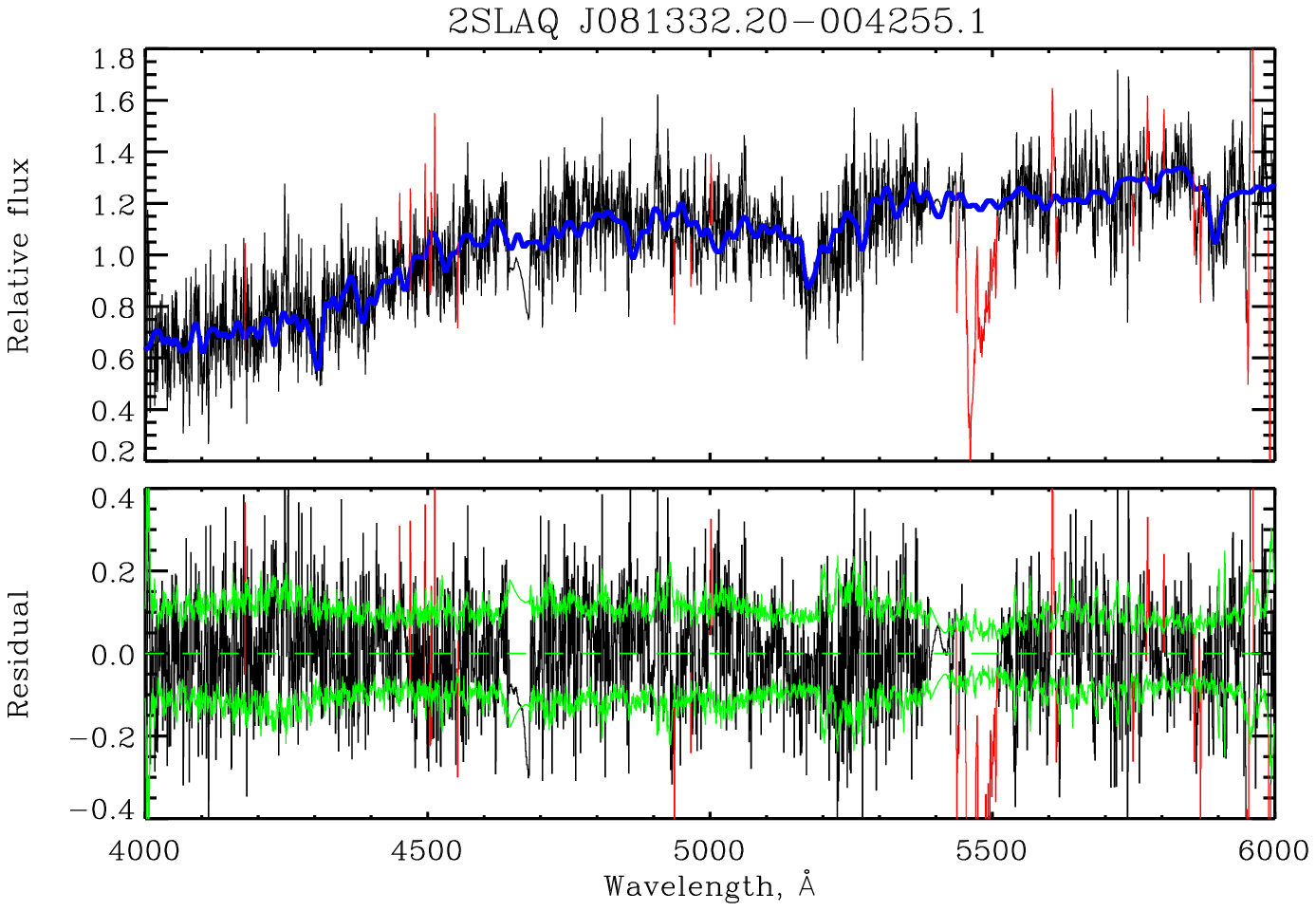,width=0.4\linewidth,clip=} &
\epsfig{file=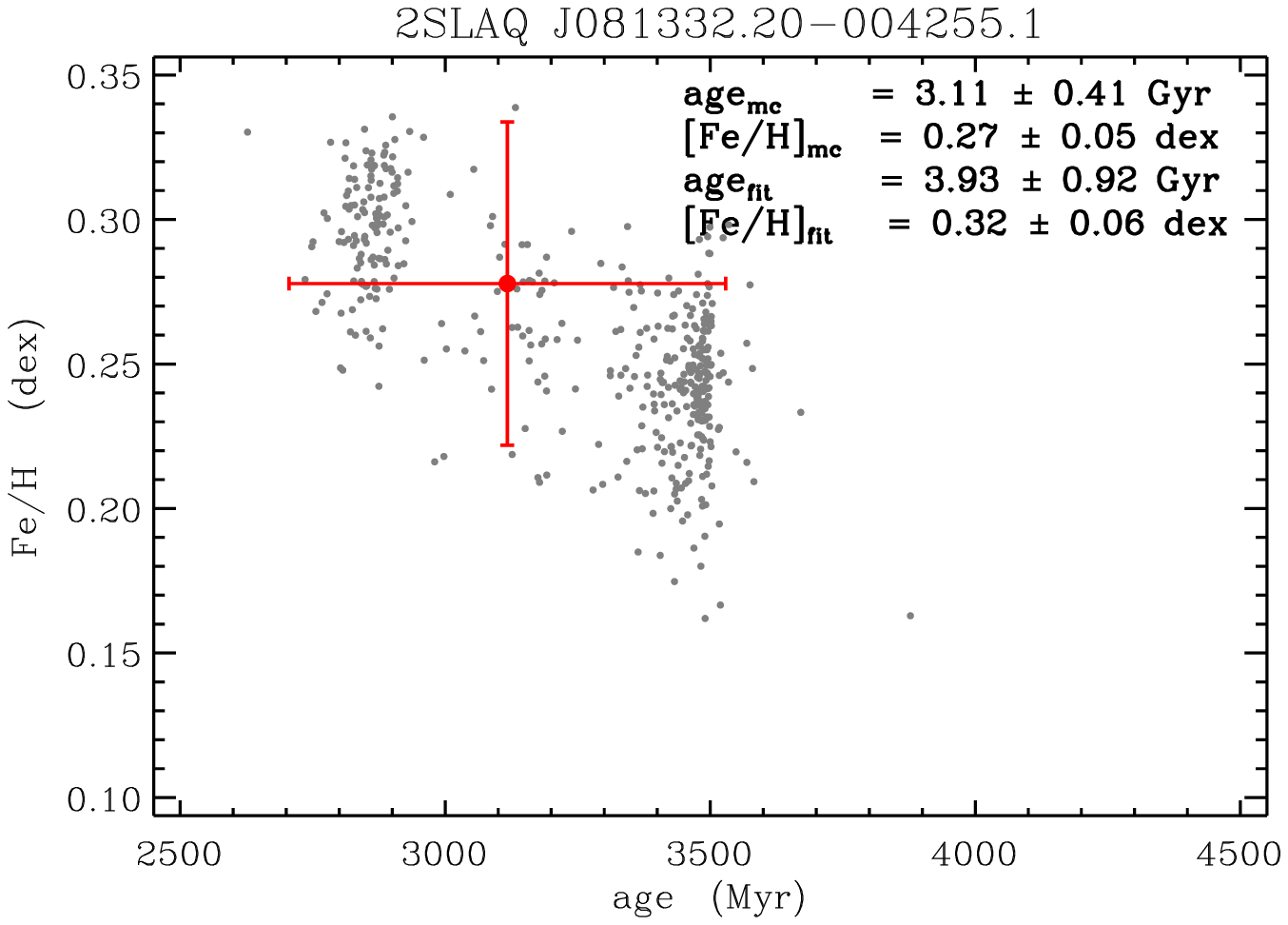,width=0.4\linewidth,clip=}
\end{tabular}
\caption{All the fits performed with \textsc{ulyss}. The left-hand panel shows individual spectral fitting. Galaxy spectra are in black and best fitted in blue line. The red regions were excluded and masked in the fit. The green lines in the residuals from the fit are the estimated 1--$\sigma$ deviation. The right-hand panel illustrates the results from the 500 Monte Carlo simulations. The red crosses indicate the mean values and standard deviations of the distributions of the age and metallicity. Some obvious outliers detached from the main distribution were excluded for the mean value measurements. The age (age$\rm_{mc}$) and metallicity ([Fe/H]$\rm_{mc}$) values from the simulations indicated are compared with those provided by single fits (age$\rm_{fit}$ and [Fe/H]$\rm_{fit}$ written in the legend).}
\label{plotappendix:saltspec}
\end{figure*}

\addtocounter{figure}{-1}
\begin{figure*}
\centering
\begin{tabular}{cc}
\epsfig{file=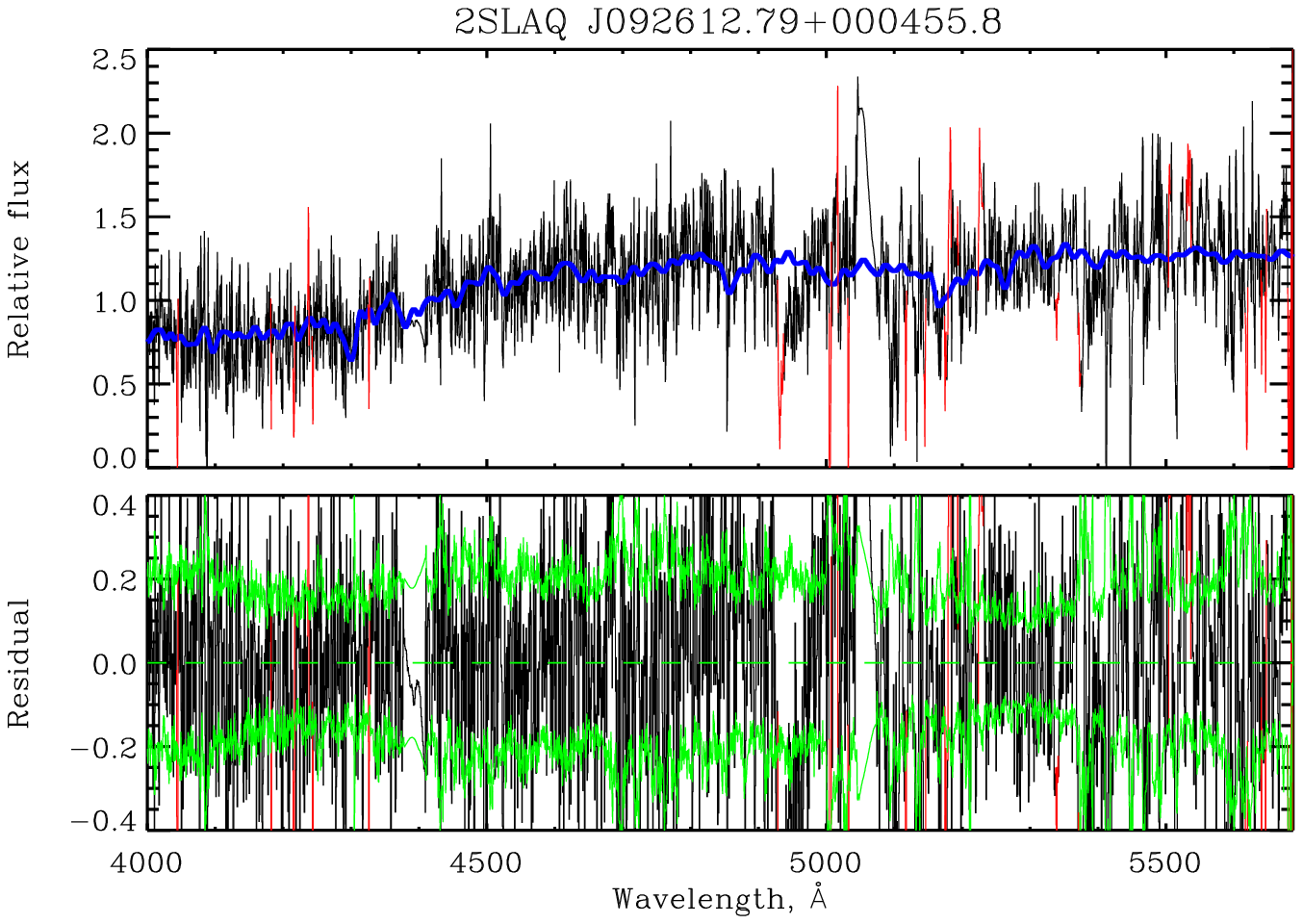,width=0.4\linewidth,clip=} &
\epsfig{file=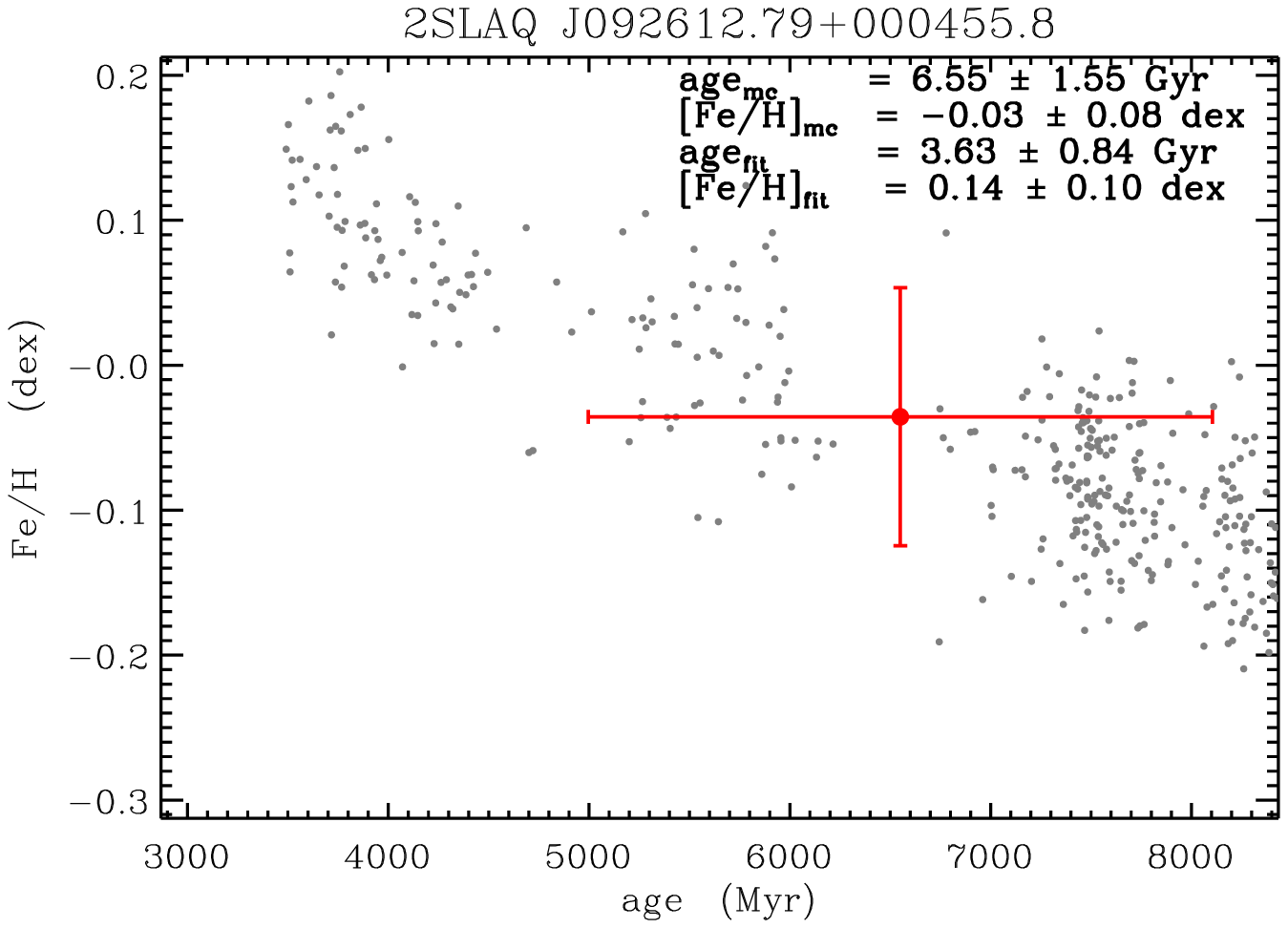,width=0.4\linewidth,clip=}\\
\epsfig{file=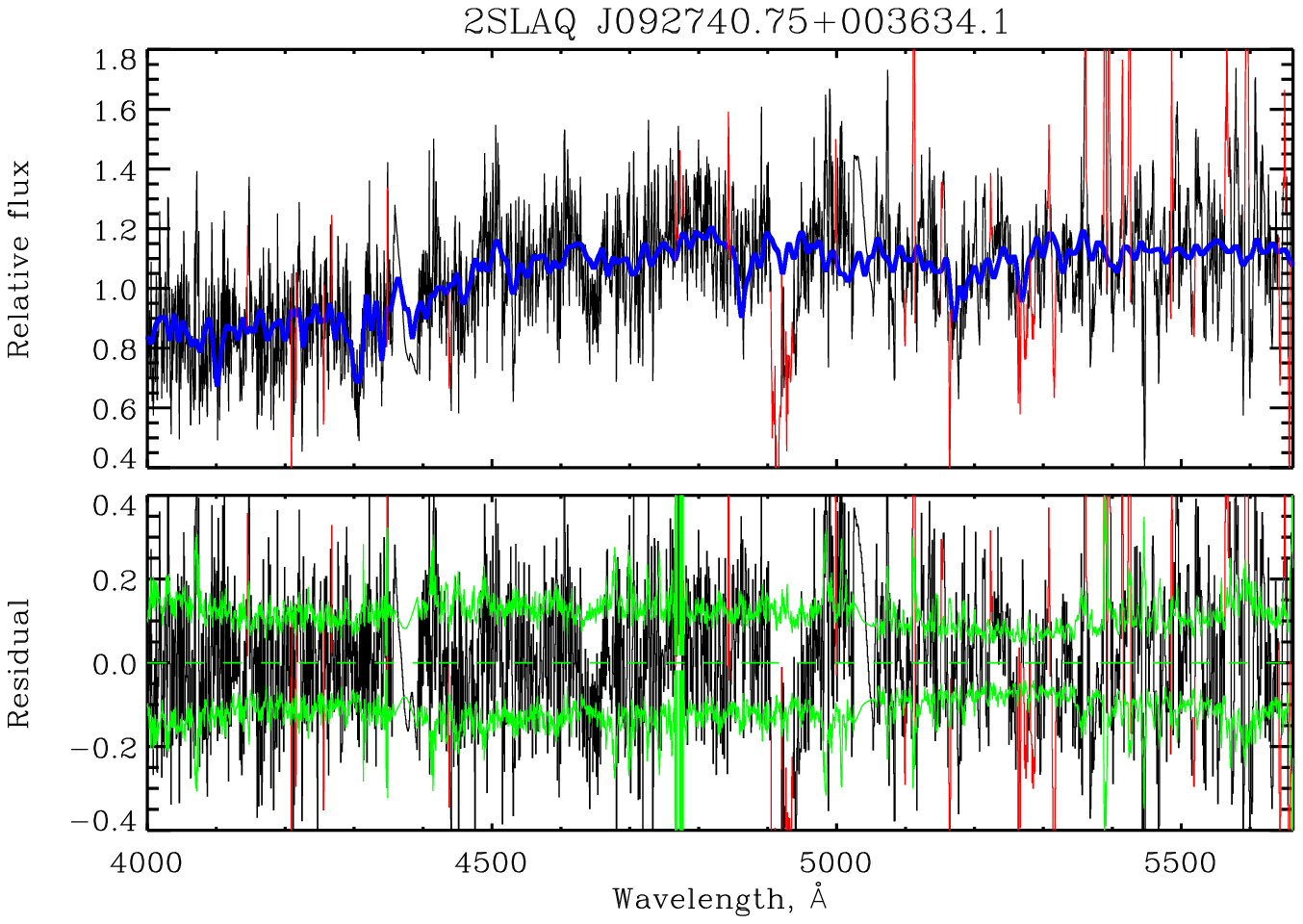,width=0.4\linewidth,clip=} &
\epsfig{file=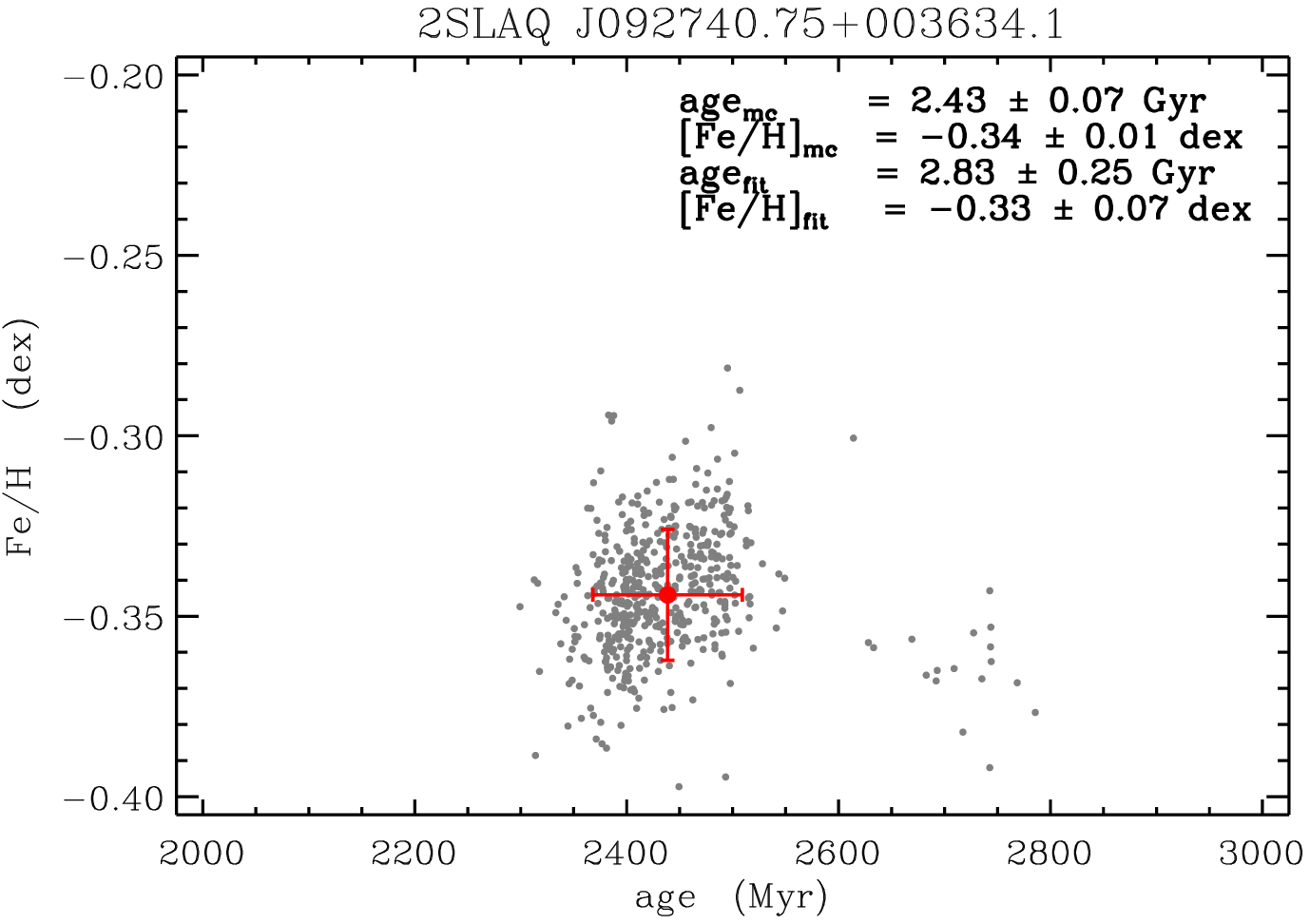,width=0.4\linewidth,clip=}\\
\epsfig{file=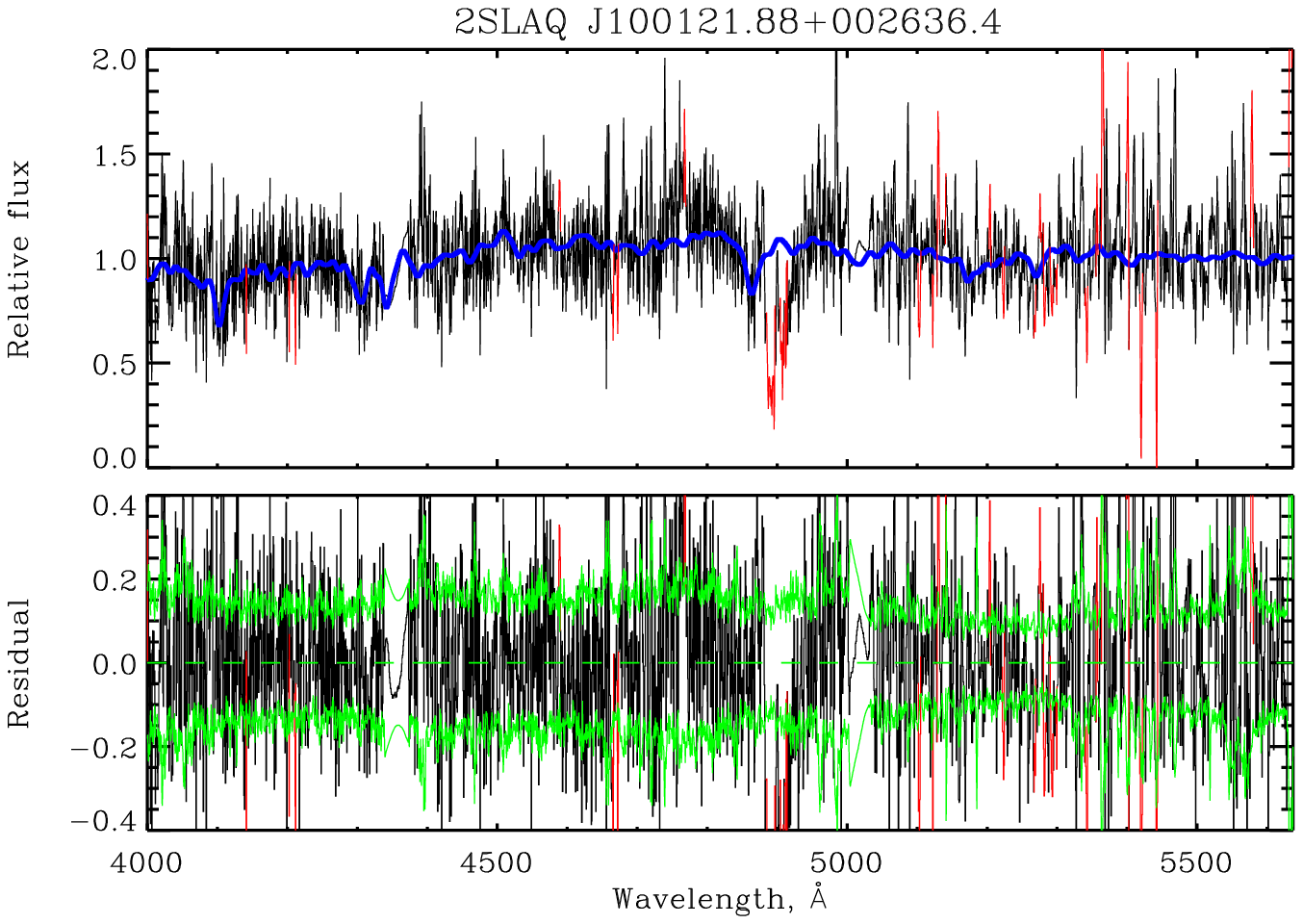,width=0.4\linewidth,clip=} &
\epsfig{file=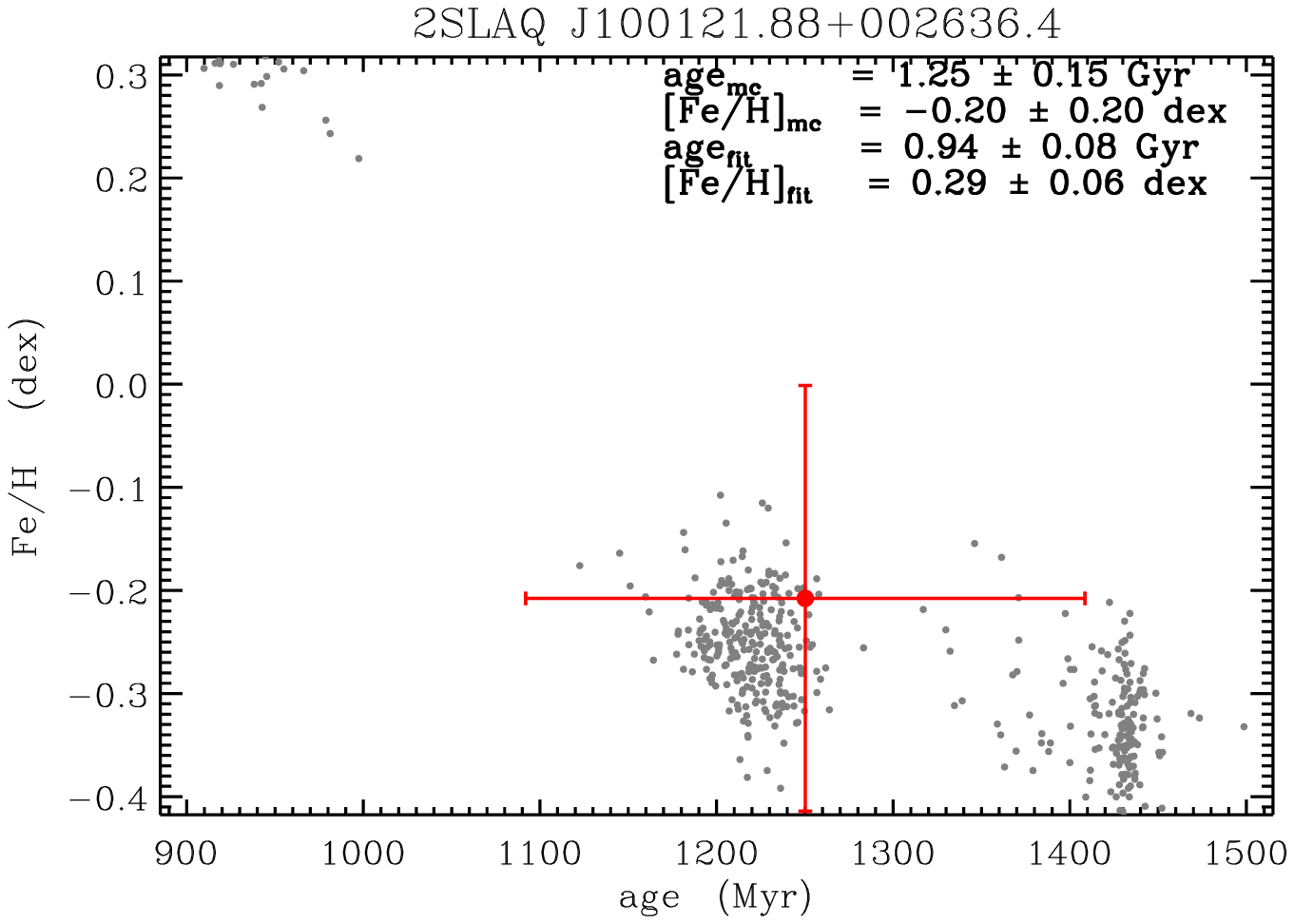,width=0.4\linewidth,clip=}\\
\epsfig{file=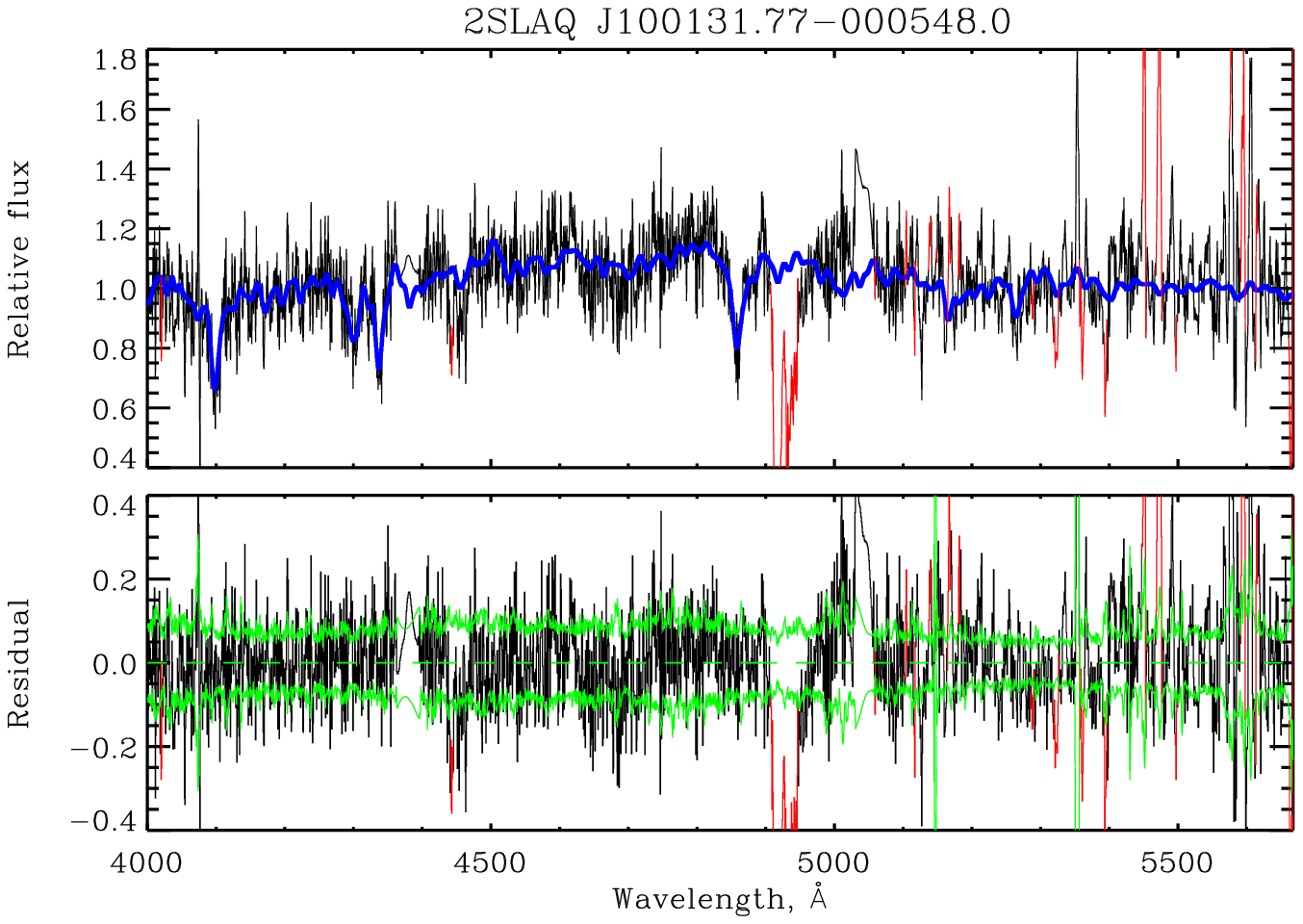,width=0.4\linewidth,clip=} &
\epsfig{file=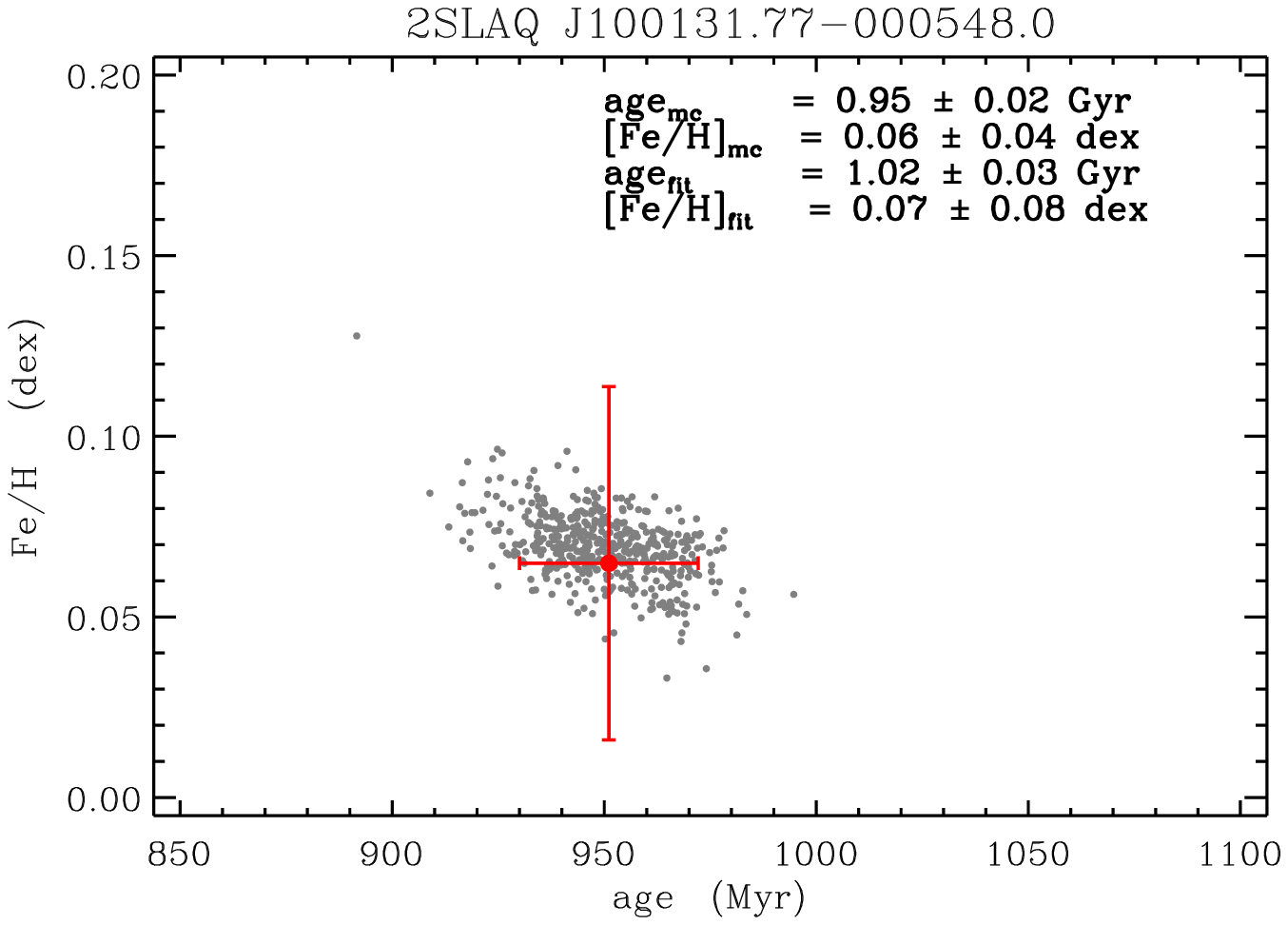,width=0.4\linewidth,clip=}
\end{tabular}
\caption{continued.}
\label{saltspec}
\end{figure*}

\addtocounter{figure}{-1}
\begin{figure*}
\centering
\begin{tabular}{cc}
\epsfig{file=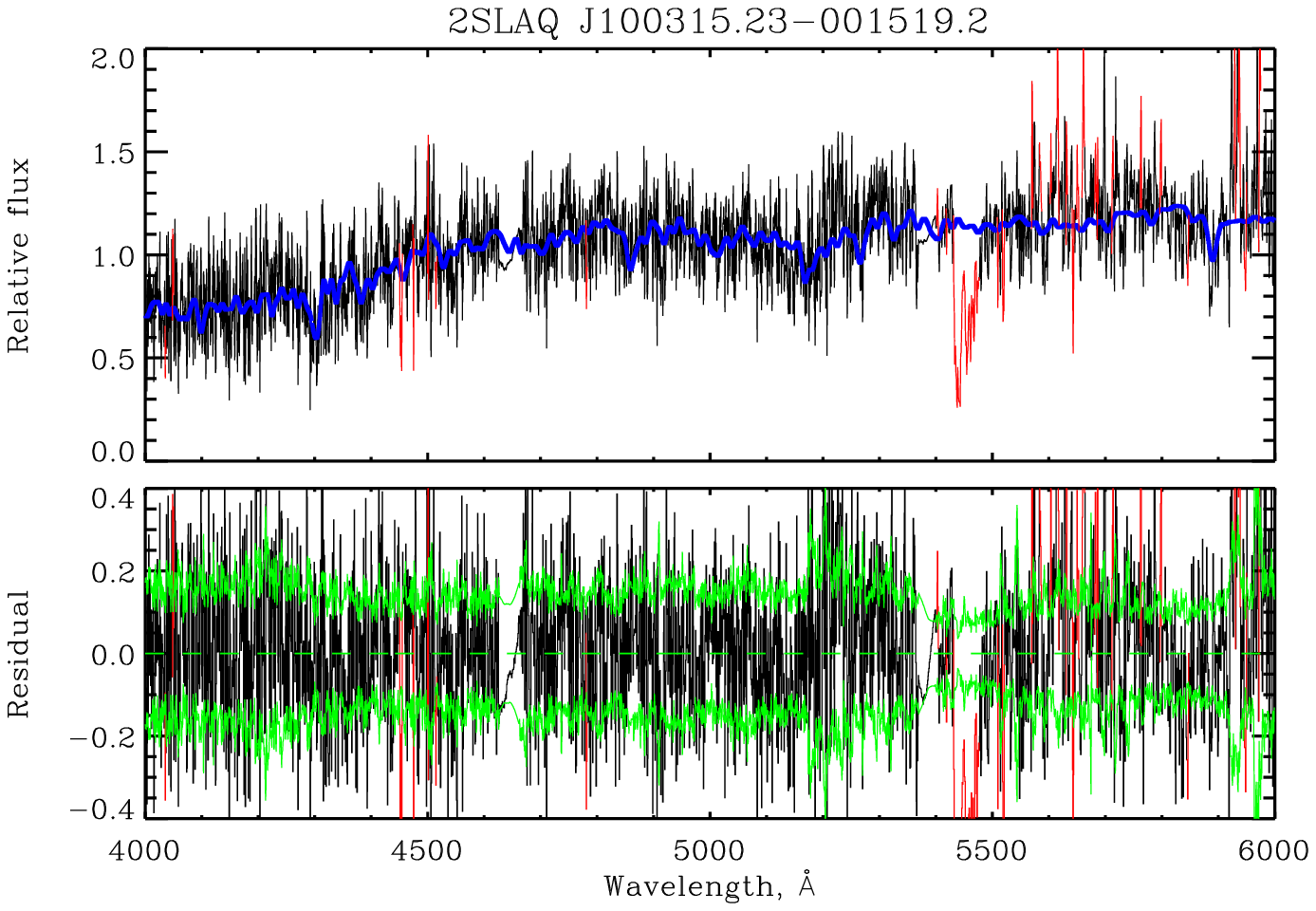,width=0.4\linewidth,clip=} &
\epsfig{file=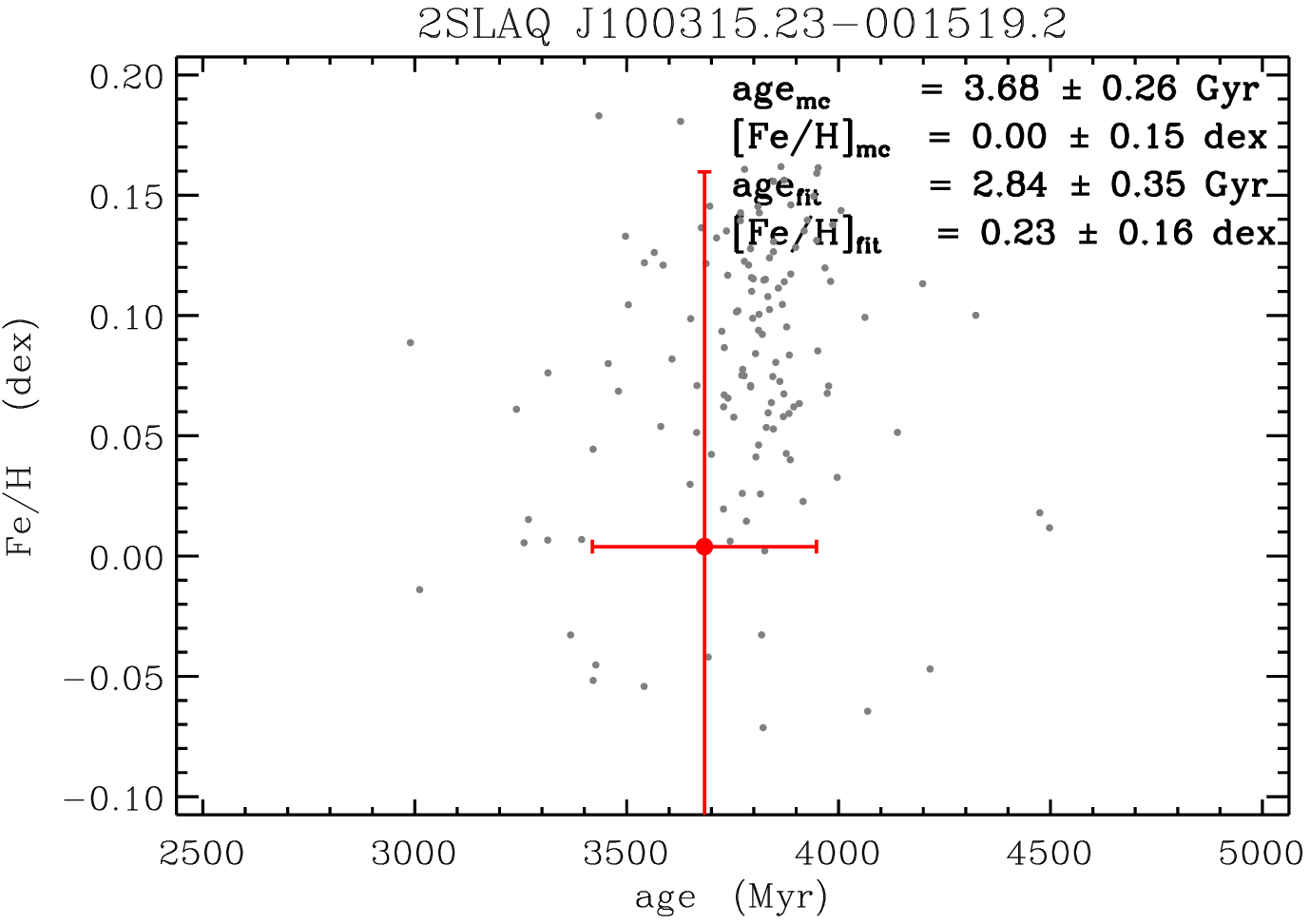,width=0.4\linewidth,clip=}\\
\epsfig{file=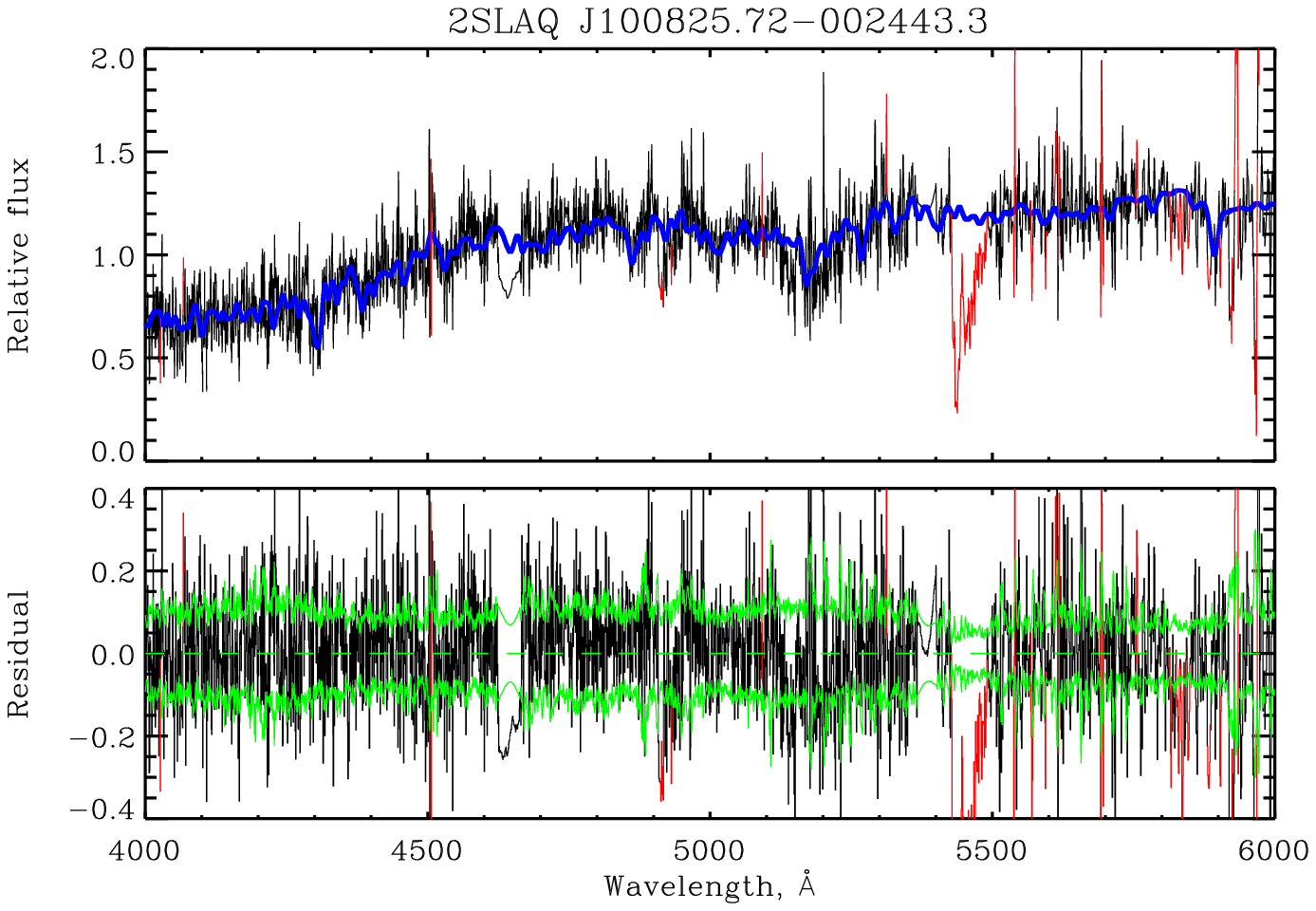,width=0.4\linewidth,clip=} &
\epsfig{file=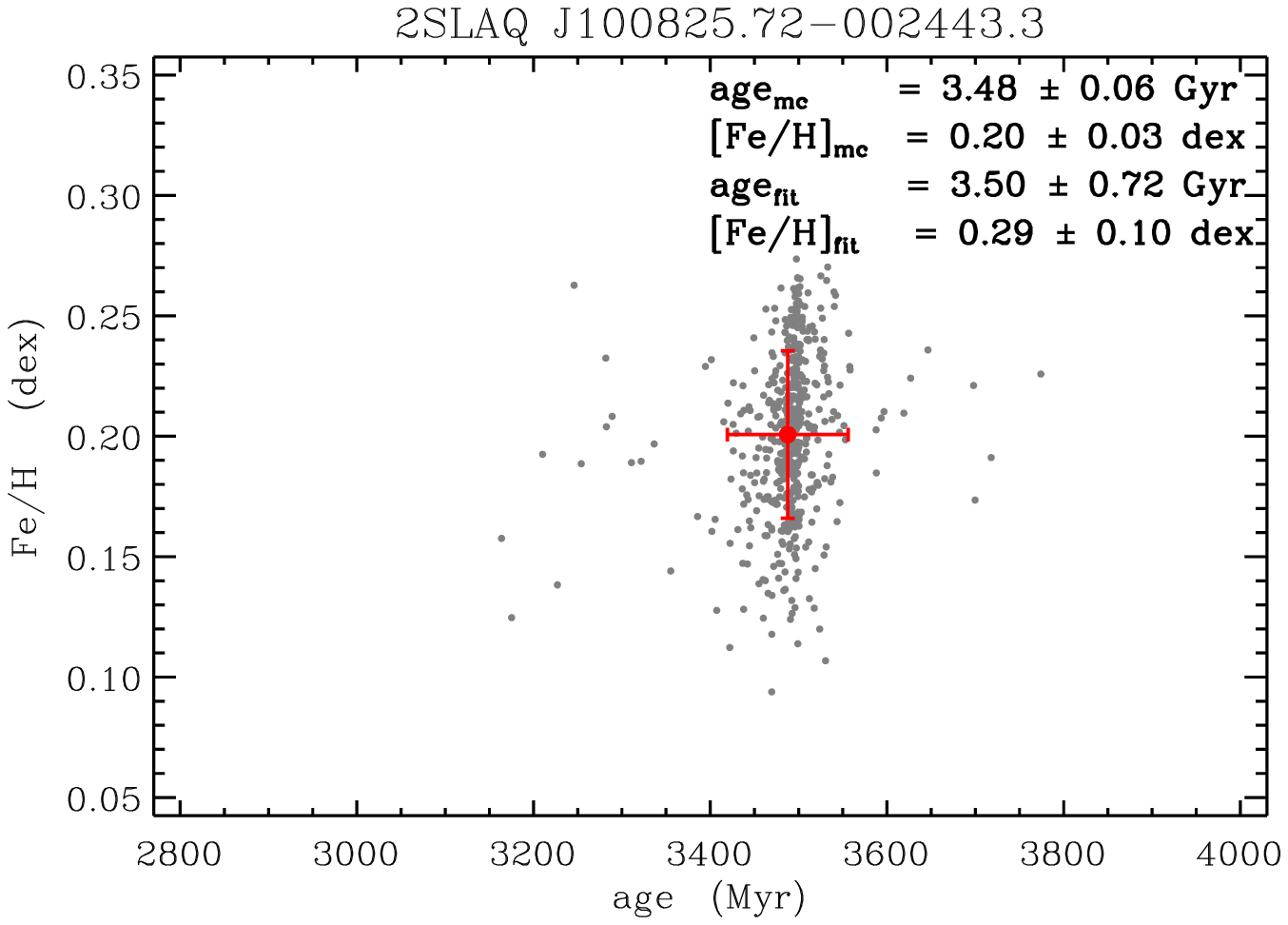,width=0.4\linewidth,clip=}\\
\epsfig{file=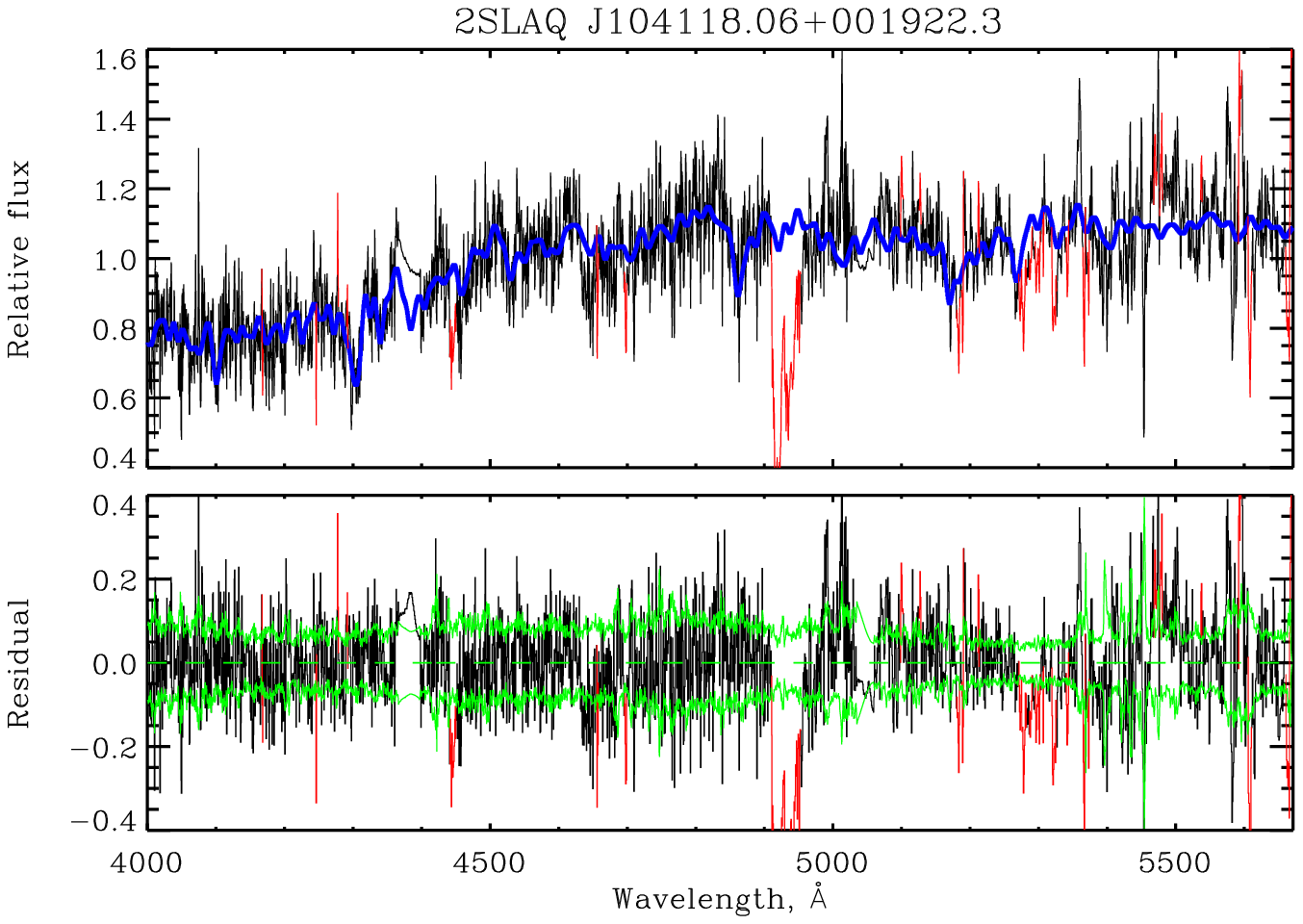,width=0.4\linewidth,clip=} &
\epsfig{file=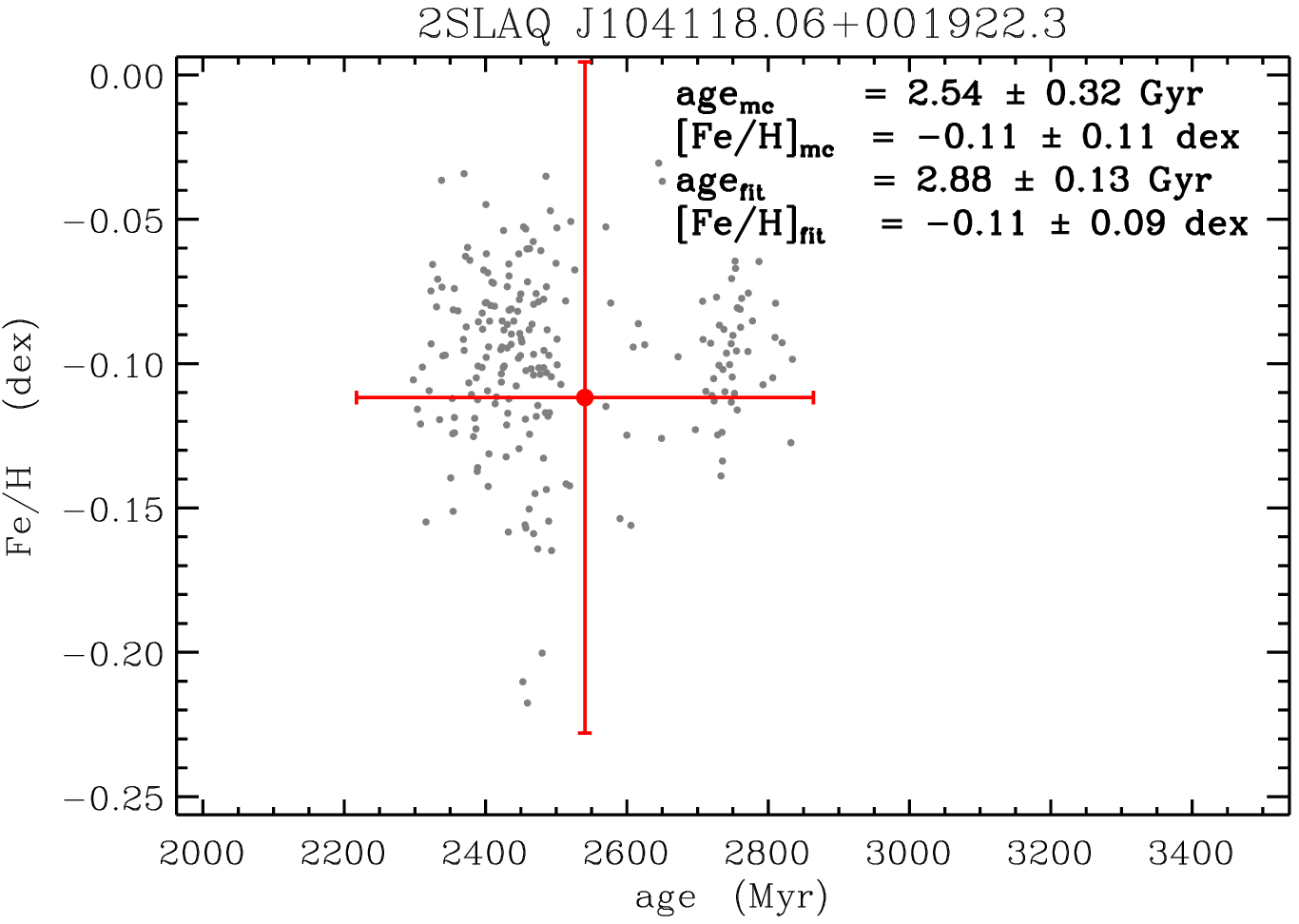,width=0.4\linewidth,clip=}\\
\epsfig{file=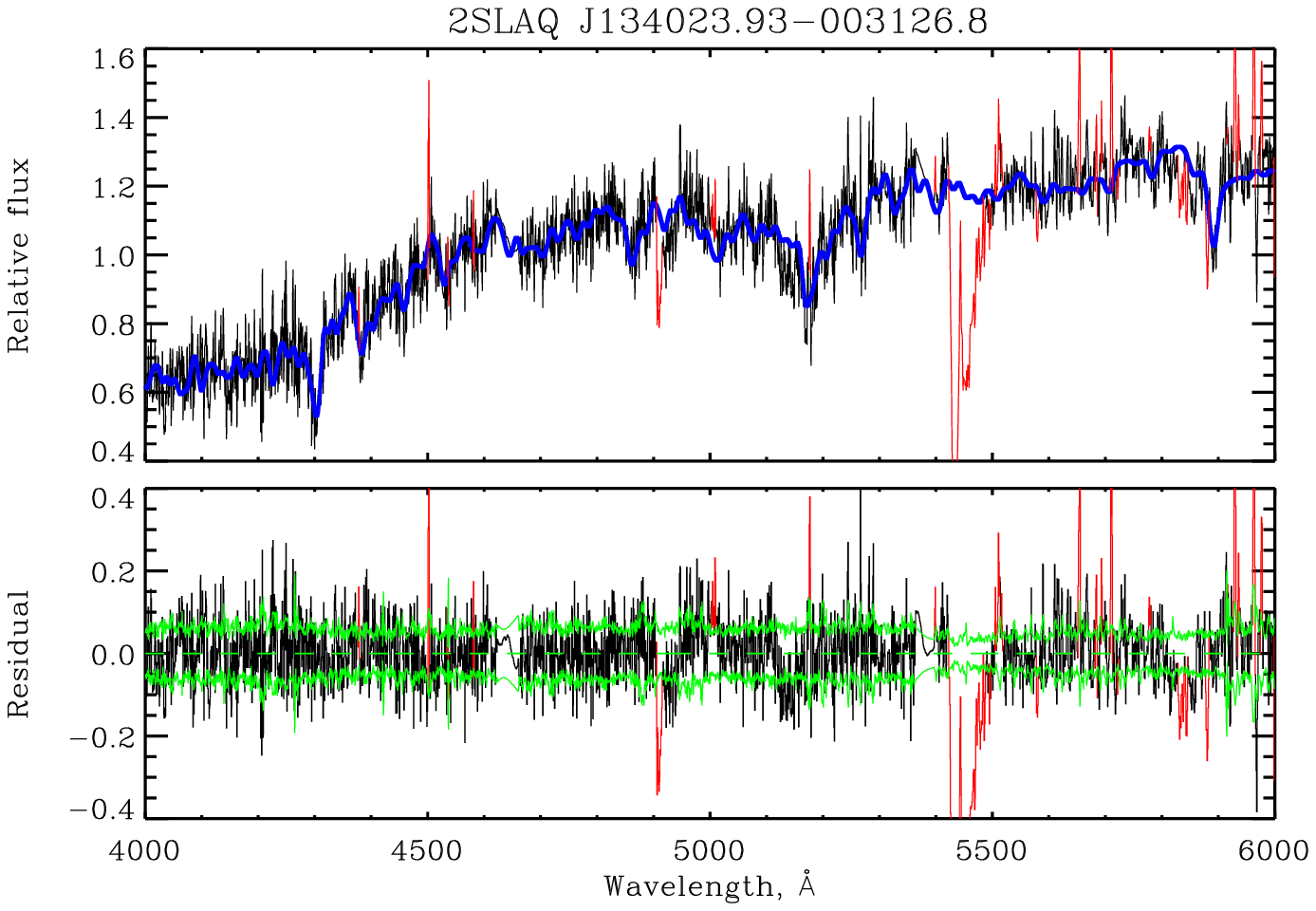,width=0.4\linewidth,clip=} &
\epsfig{file=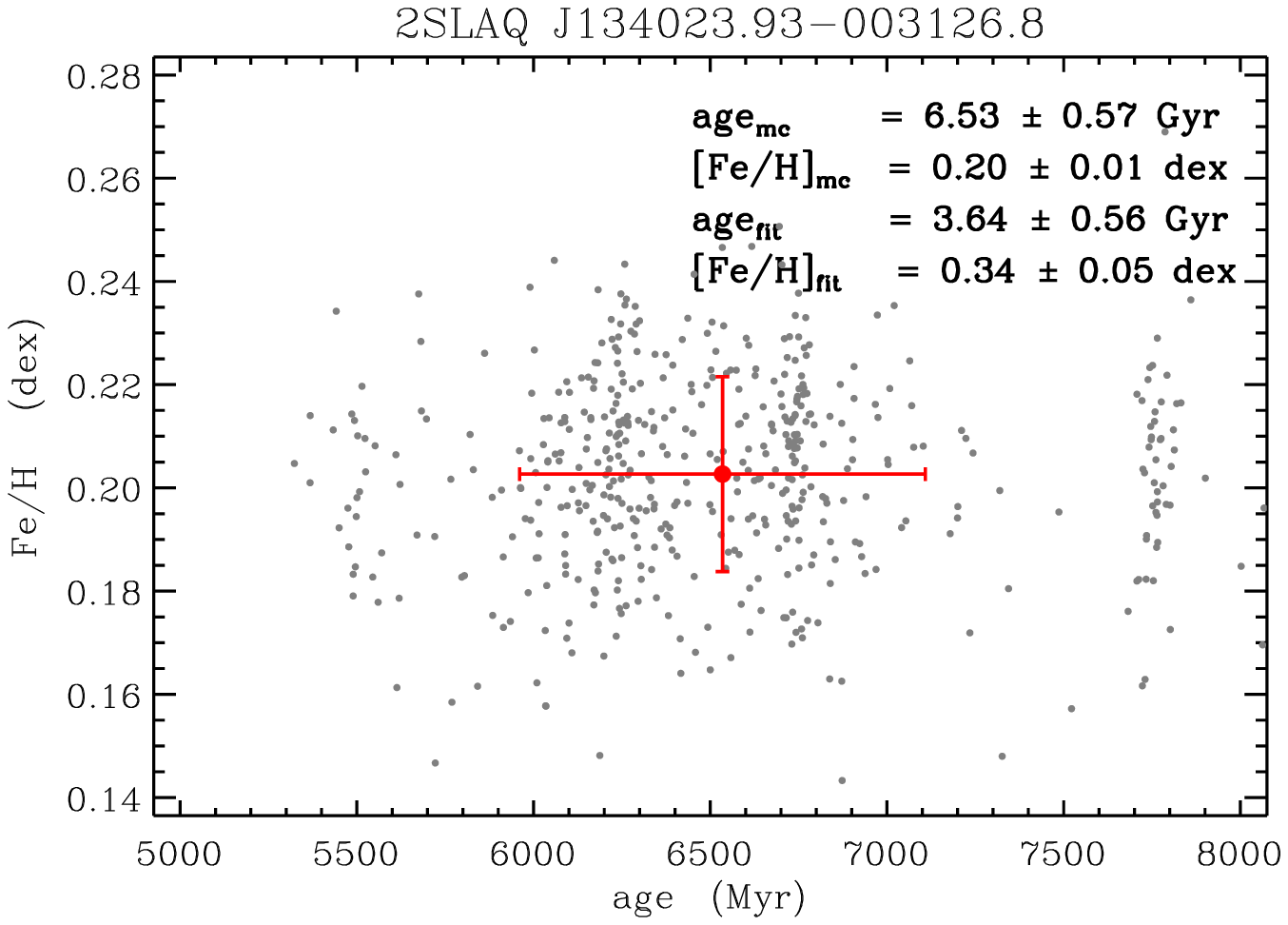,width=0.4\linewidth,clip=}
\end{tabular}
\caption{continued.}
\label{saltspec}
\end{figure*}

\addtocounter{figure}{-1}
\begin{figure*}
\centering
\begin{tabular}{cc}
\epsfig{file=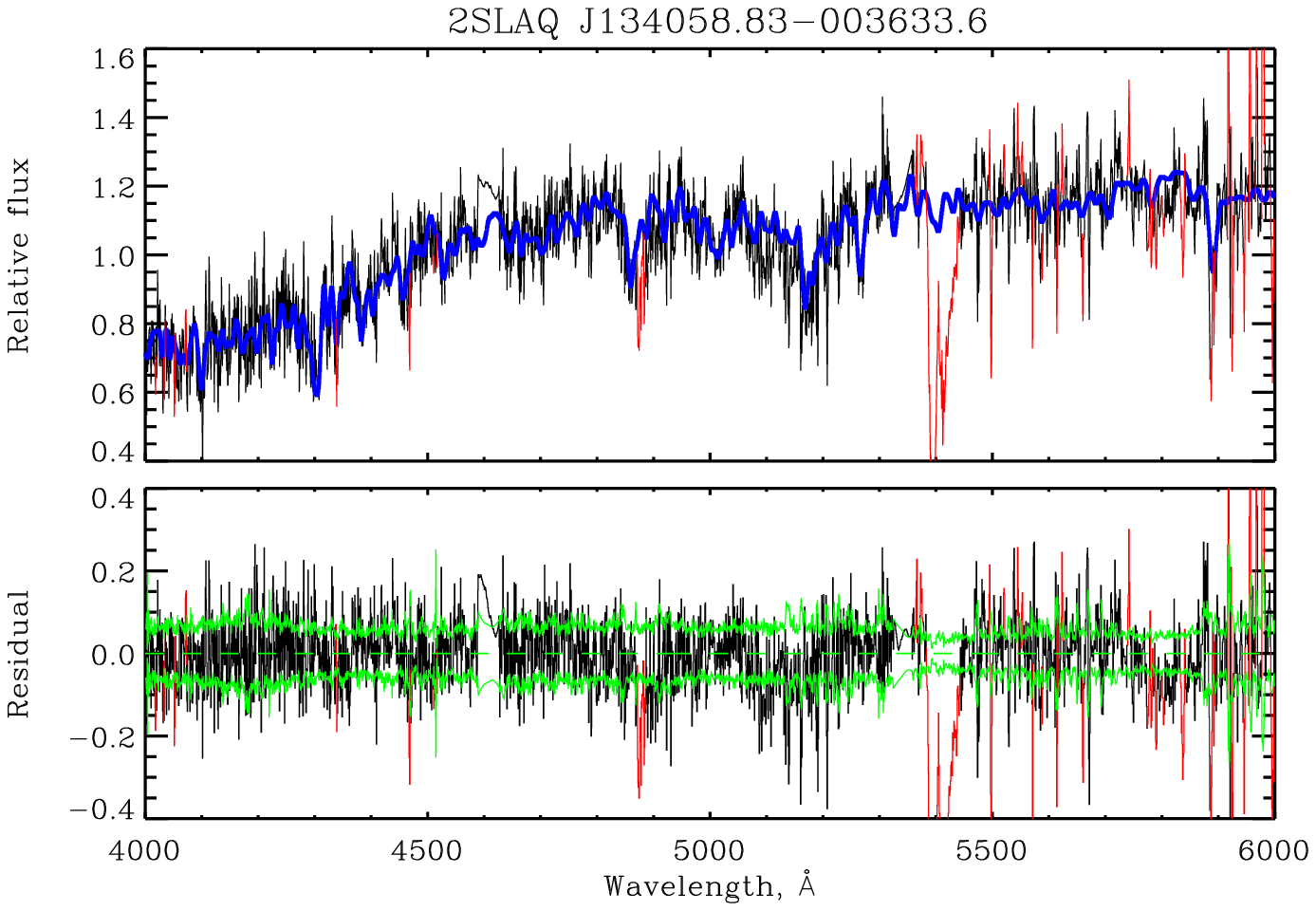,width=0.4\linewidth,clip=} &
\epsfig{file=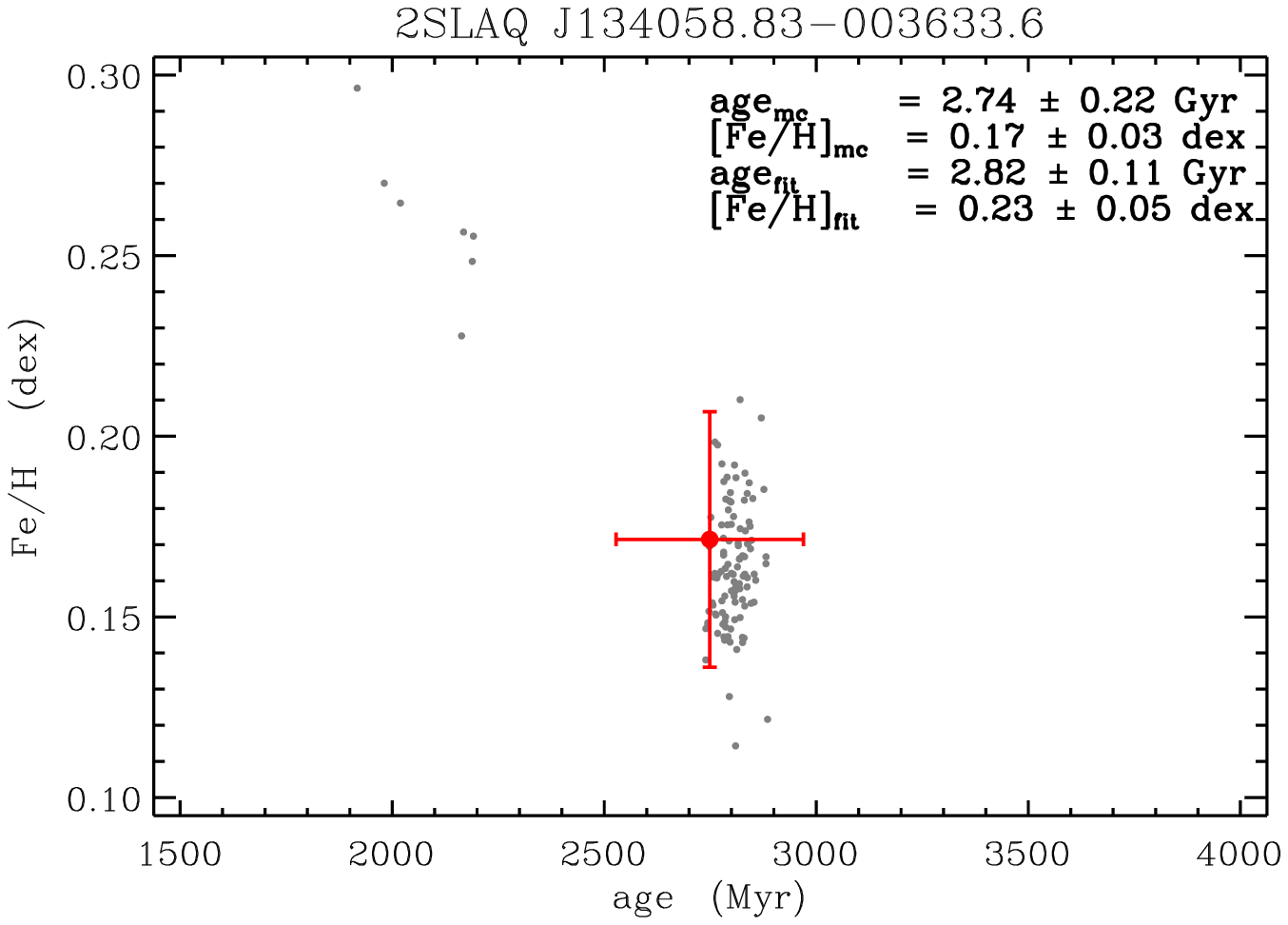,width=0.4\linewidth,clip=}\\
\epsfig{file=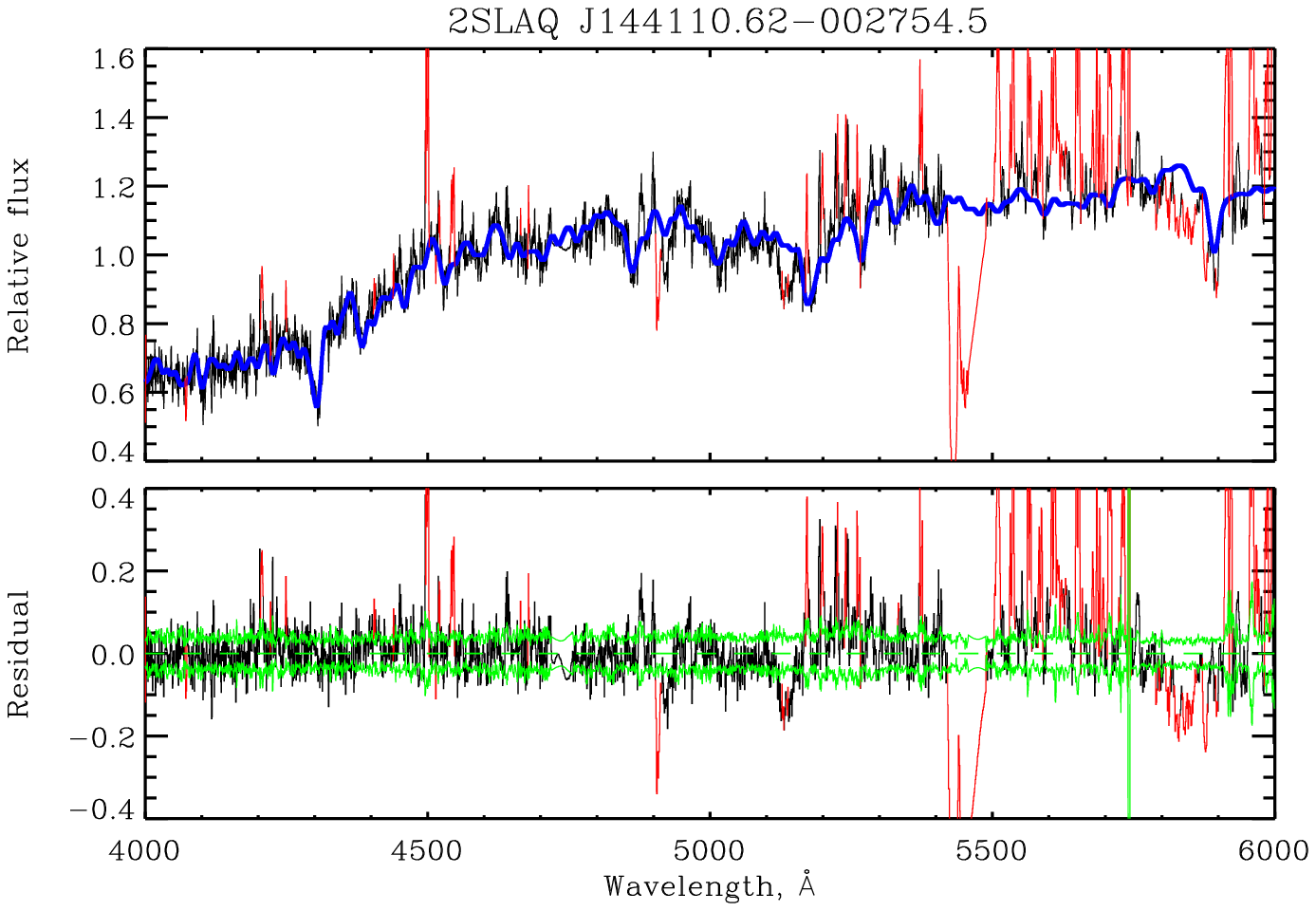,width=0.4\linewidth,clip=} &
\epsfig{file=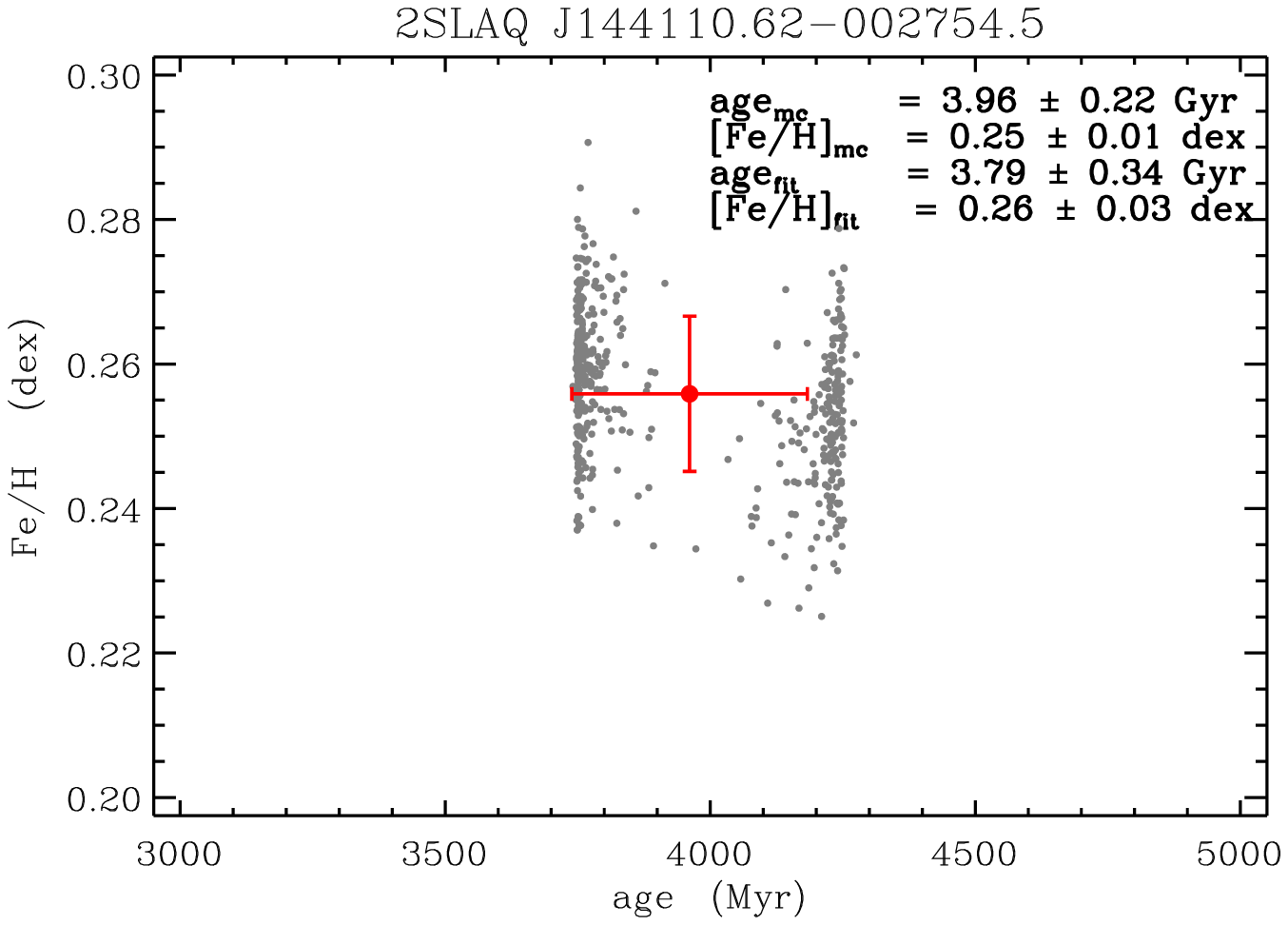,width=0.4\linewidth,clip=}\\
\epsfig{file=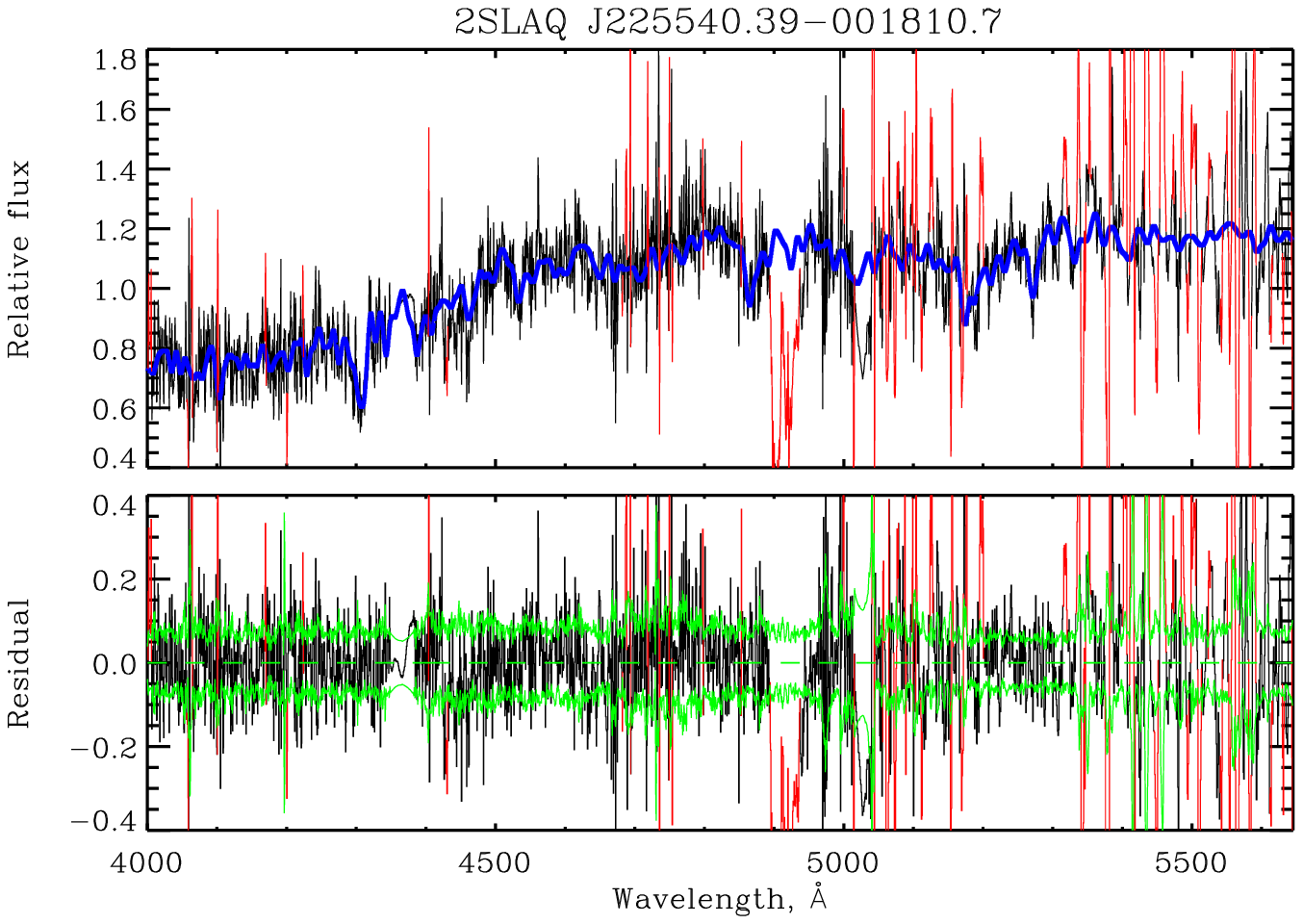,width=0.4\linewidth,clip=} &
\epsfig{file=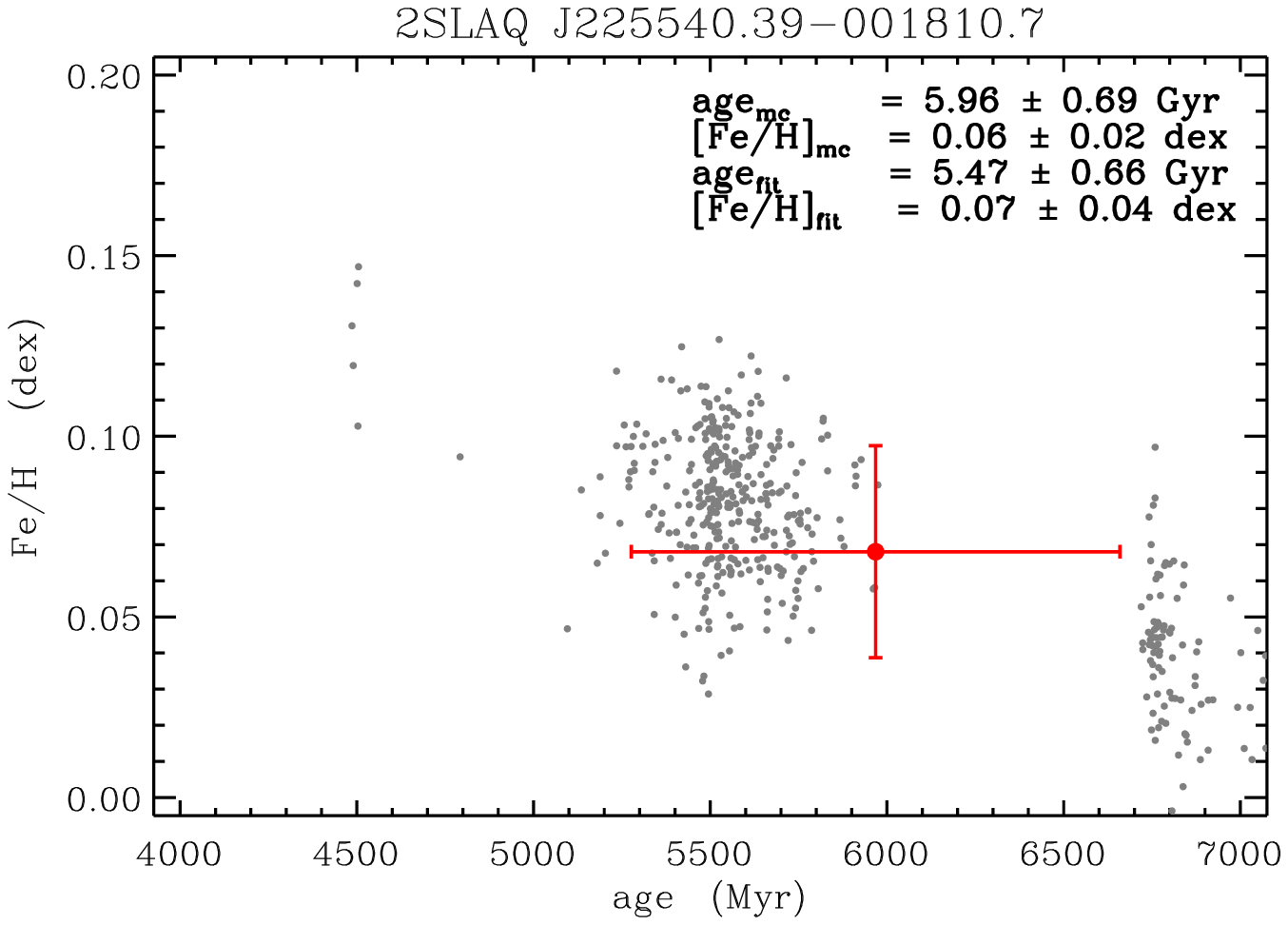,width=0.4\linewidth,clip=}\\
\epsfig{file=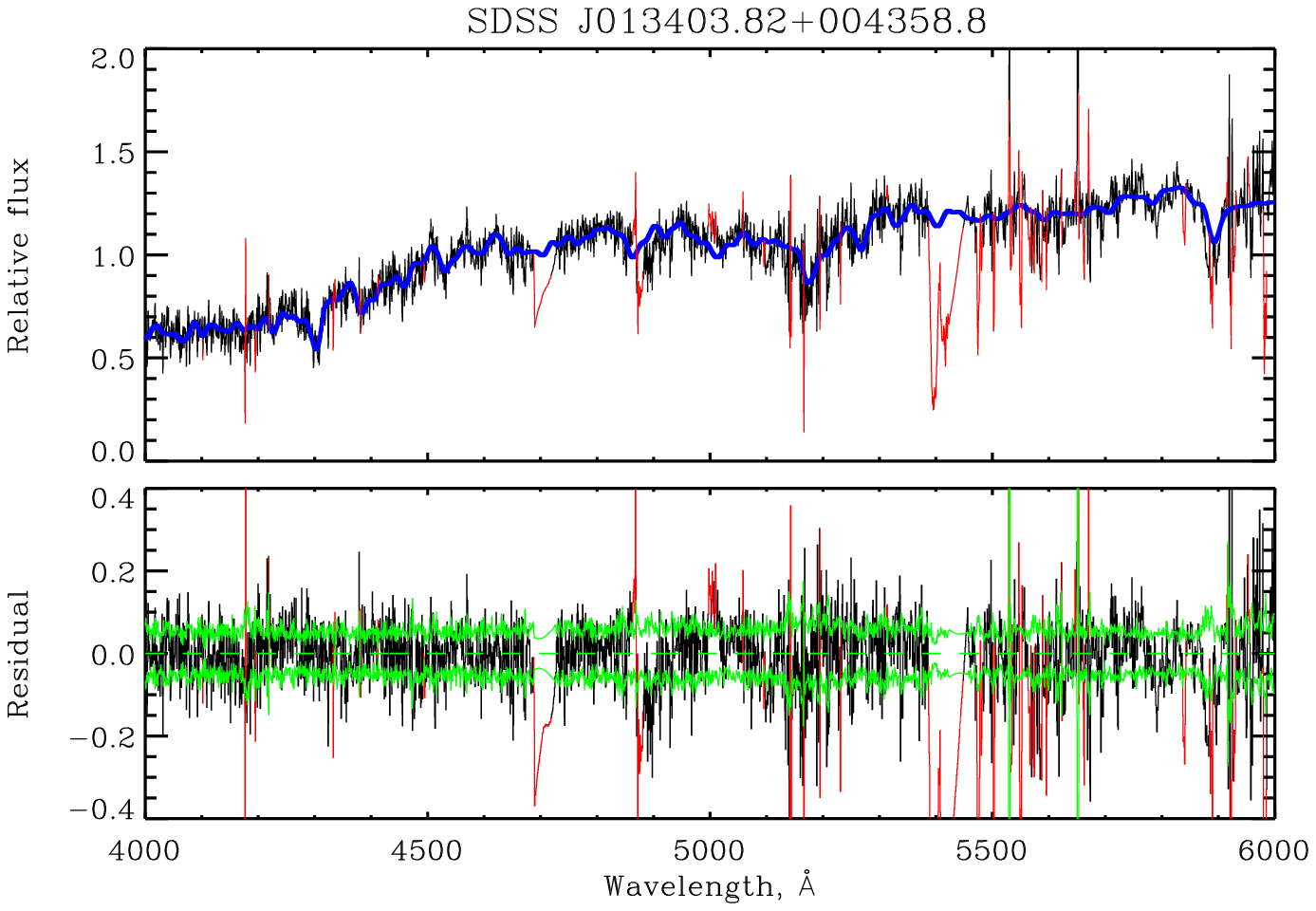,width=0.4\linewidth,clip=} &
\epsfig{file=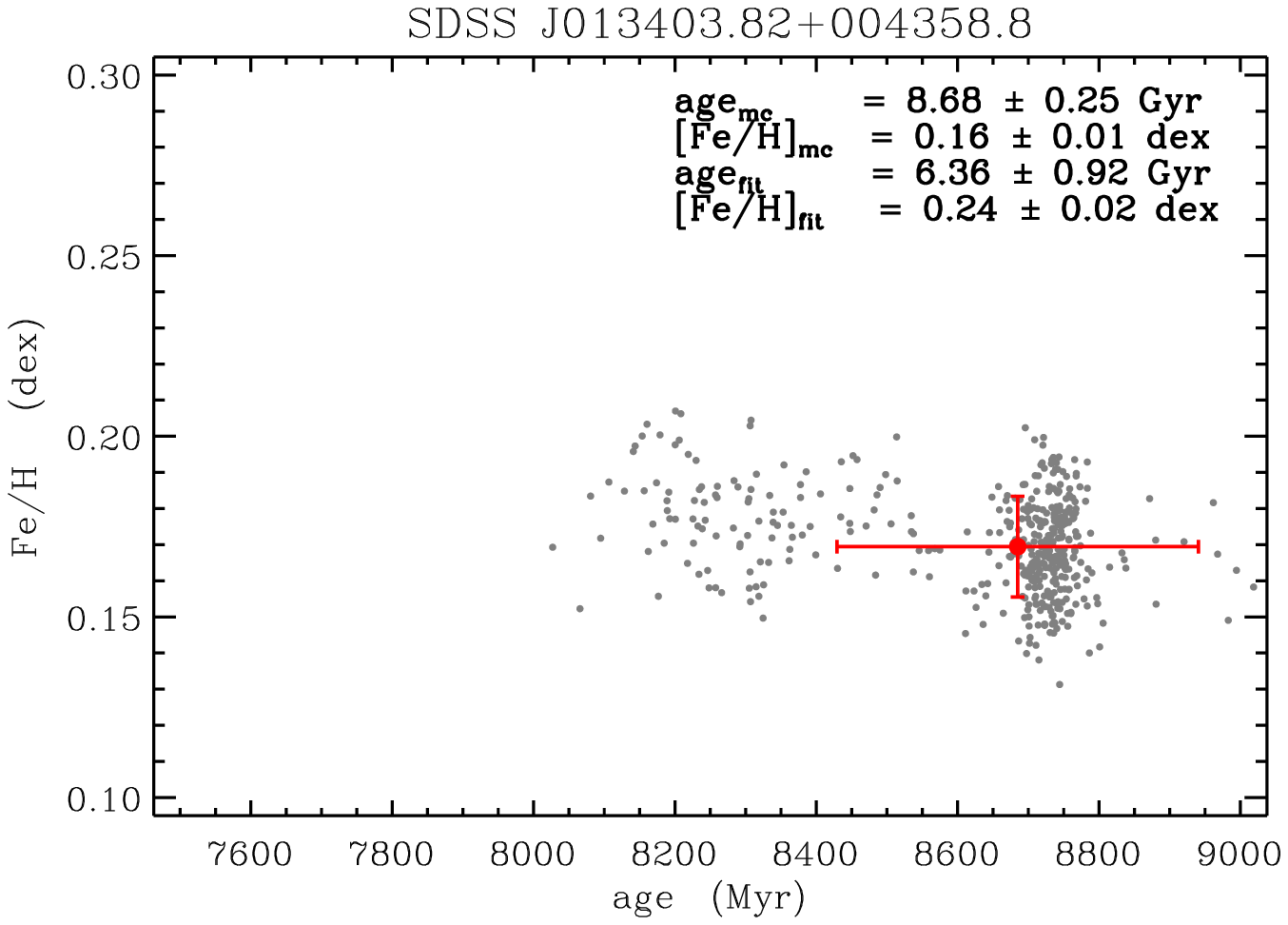,width=0.4\linewidth,clip=}
\end{tabular}
\caption{continued.}
\label{saltspec}
\end{figure*}

\bsp	
\label{lastpage}
\end{document}